# 定量分析调控 HIV 潜伏与激活的动力学机制[*]

熊瑞琪[**]　苏洋[**]　敖平[***]

（上海大学物理系，定量生命科学国际研究中心，上海 200444）

**摘要**　**目的**　治疗艾滋病最大的障碍在于无法根除人类免疫缺陷病毒（HIV）潜伏于人体细胞所形成的病毒存储库。构建描述病毒存储库建立的分子机制的动力学模型需考虑生物体内的噪声环境和多重影响因素，本文通过一种全新的动力学结构分解方法将随机微分方程的确定性部分与随机性噪声分开，从而在仅需分析常微分方程不动点的情况下即可判断不同药物靶点的作用效果。**方法**　使用连续的随机微分方程构建了 HIV 转录过程的动力学模型，简化了描述系统所需方程的维度，增大了模型的可探索空间，在此基础上，通过计算得到的势能函数和概率分布函数直观表示病毒潜伏与激活的不同表达状态以及它们之间的关系。**结果**　定量分析了不同动力学参数对系统稳态和势函数的影响程度，分别得到了系统处于双稳态和单稳态时的参数范围，并将不同因素对动力系统分岔的影响程度与生物学实验结果对比，验证了本工作的理论基础。**结论**　本文突破了以往离散、随机的方法，可以通过常微分方程定量分析 HIV 转录调控的动力学机制，有利于推广到处理高维情况，进一步研究艾滋病在生物体内的发生发展，从而指导设计实验寻找临床上的治疗方案。

**关键词**　HIV，常微分方程，动力学结构分解，双稳态，分子开关，势函数
**中图分类号**　Q3，Q615，Q31

人类免疫缺陷病毒（human immunodeficiency virus，HIV）是获得性免疫缺陷综合征（acquired immune deficiency syndrome，AIDS）即艾滋病的病原体。艾滋病是一种广泛传播的流行病，截至2020年，全球约有3770万HIV感染者（https://aidsinfo.unaids.org）。目前治疗HIV感染的主要方法为联合使用多种抗HIV药物的高效抗逆转录病毒疗法（highly active antiretroviral therapy，HAART），该疗法能有效抑制艾滋病病毒的复制，将血液中的病毒载量降低至检测不到的水平，进而控制病情发展[1]。但处于潜伏态的病毒可以逃逸机体的免疫清除和抗病毒治疗，形成体内病毒潜伏存储库，在停止抗病毒治疗或治疗失败后，潜伏态病毒将会重新激活，导致病毒数量的迅速反弹，因此潜伏存储库是彻底治愈艾滋病的最大障碍。

为实现联合国艾滋病规划署到2030年在全球消除艾滋病毒的目标，目前关于抗HIV治疗的研究旨在清除病毒潜伏存储库，包括病毒转录编辑、基因编辑、激活并杀死（shock and kill）、阻断和锁定（block and lock）、干细胞移植和基因特异性转录激活等策略[2]。其中研究最多的是"激活并杀死"策略，即使用潜伏反转剂（latency reversing agents，LRAs）重新激活处于潜伏状态的病毒，使得受感染的细胞被病毒本身或患者的免疫系统杀死，从而清除潜在病毒存储库[3]。另外，"阻断和锁定"策略的目的则是创造一个永久性潜伏期，即通过病毒的转录和转录后基因沉默，产生一个永久的非生产性感染状态，从而抑制病毒的再次激活[4]。了解HIV在潜伏状态与激活状态之间转换机制的理论基础对"激活并杀死"和"阻断和锁定"这两种治疗策略的发展都有重要的指导意义。

在现代信息时代中，开关式结构是所有体系结构的重要组成部分。事实上，详细的分析已经表明，一个复杂网络的响应通常由各种开关[5]来处理，并且遗传网络被证明具有计算能力[6]。如今，生物开关的研究在细胞周期等过程中发挥了重要的作用，如λ噬菌体生命周期的基因开关[7]。同样地，对于目前全球大流行的新型冠状病毒肺炎（corona virus disease 2019，COVID-19）而言，潜伏态与激活态相互转换的分子机制也可以类似看作一个开关，不同的环境信号将导致其最终的发育途径。并且开关确实已经是生物过程的基本要素之一，也是生物学实验和理论研究的范例[8]。

HIV入侵宿主细胞后整合进宿主基因组形成原病毒（provirus），原病毒的复制受到转录起始和转录延伸两个过程的控制，其中转录起始过程主要受到宿主细胞因子的调控，而在转录延伸过程中病毒反式激活蛋白（trans-activator of transcription，Tat）构成的正反馈机制可以独立于宿主细胞状态调控原病毒在快速复制的激活态与转录沉默的潜伏态之间转换[9]，这样的转换就可以看作一个生物系统中的开关。目前已有理论[10]分析Tat在促进HIV转录延伸的过程中，其乙酰化速率和结合转录激活应答元件（trans-activation response，TAR）RNA的速率等因素对病毒表达状态的影响，但当前理论[11-12]认为HIV调控网络的确定性动力学不足以解释所观察到的双稳态，需要使用随机性动力学来描述此机制。然而通过离散随机方程构建的模型使变量空间的探索范围受到了一定程度的限制，使得进一步的推广存在困难，并且目前的模型中都只包含了一部分影响因素，没有在同一个模型中对比所有不同影响因素的重要程度，存在一定的局限性。

本文使用连续的随机微分方程构建了调控HIV转录的分子开关的动力学模型，将病毒的不同表达状态对应于系统的不同稳态，通过势函数和概率分布函数定量分析了艾滋病潜伏期建立的分子机制。根据动力学结构分解下随机微分方程极值点与常微分方程稳定点的对应，我们仅需分析模型中方程的确定性部分就可以判断系统处于单稳态或双稳态的不同情况，从而分析病毒表达状态。在此基础上，我们参考现有实验数据计算出了使稳态个数发生改变时的各影响因素的参数区间，并在理论上对比了不同影响因素对于潜伏态与激活态之间转换的影响程度，为"激活并杀死"与"阻断和锁定"治疗策略提供了理论基础。本研究使用的连续方法相较于过去的离散方法在很大程度上缩减了变量存储空间的大小，可以更加方便地扩大模型中可调参数的变化范围、增加变量和参数的探索空间，有利于进一步发展彻底治愈艾滋病的可行的治疗靶点与药物。

# 1 材料和方法

## 1.1 HIV 转录过程

原病毒的长末端重复序列（long terminal repeat，LTR）上包含多个细胞转录因子的结合位点，启动子和增强子序列可结合宿主细胞因子如 Sp1、NF-κB、NFAT 等从而促进转录起始，当宿主细胞处于活跃状态时，宿主细胞因子数量较高，更容易结合至 LTR 促进转录起始。但在缺少 Tat 的情况下，转录起始后，由于负延伸因子（NELF）的活性和阻止 RNA 聚合酶 II（RNA polymerase II，RNApⅡ）C 端结构域（CTD）磷酸化的 DRB 敏感诱导因子（DSIF）造成的阻滞，RNApⅡ仅可在 DNA 上移动一小段距离，生成少量且不完整的转录产物。病毒蛋白 Tat 经过不同位点乙酰化以及与 TAR RNA 结合的正反馈回路，可以将转录延伸所需的细胞因子如阳性转录延伸因子 b（P-TEFb）等招募至 RNApⅡ形成复合物，解除转录阻滞，促进转录延伸，生成大量且完整的转录产物（图 1a）[9, 13]。

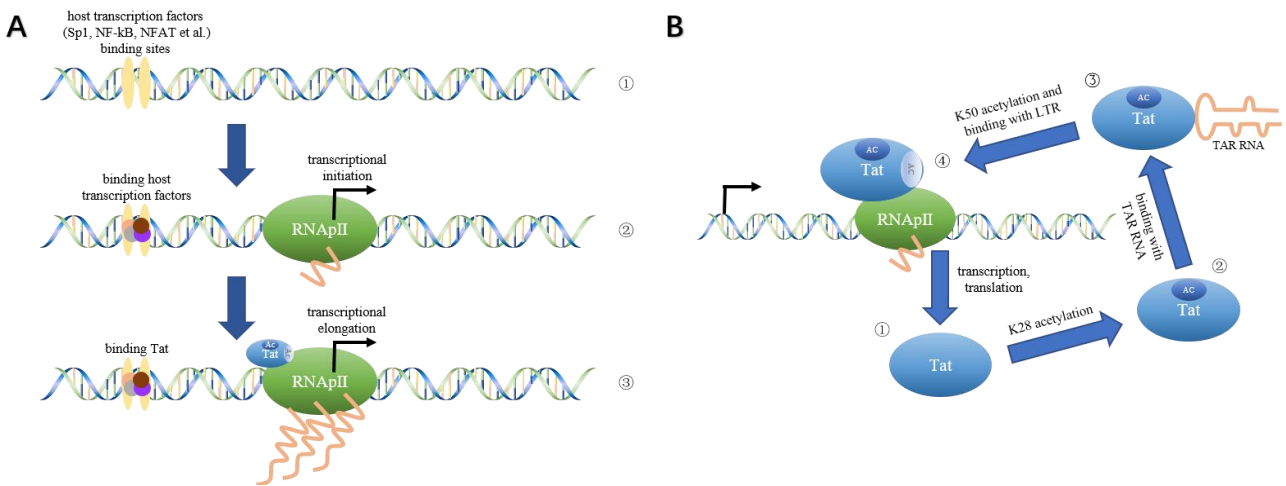

**Fig. 1 Transcriptional initiation and elongation on LTR (a) and the reaction process of Tat promoting transcriptional elongation (b)**

Tat 促进转录延伸的正反馈机制如图 1b 所示。首先，由赖氨酸乙酰转移酶（KAT）介导 Tat K28 位点的乙酰化，改变蛋白质构象，使其与由细胞周期蛋白 T1（CycT1）和细胞周期依赖性激酶 9（CDK9）组成的 P-TEFb 亲和力增强，形成复合物，并结合至 TAR RNA；然后，KAT 介导 Tat K50 位点的乙酰化，使复合物与 TAR RNA 分离，作用于 LTR 上，其中 CDK9 过磷酸化 RNApⅡ CTD，还能磷酸化 DSIF 和 NELF，使其对延伸过程的作用由负向正转变，从而可以解除 RNApⅡ受到的阻滞作用，促进转录延伸[14]。HIV LTR 上发生的转录调控相关化学反应的整体流程图如图 2 所示。

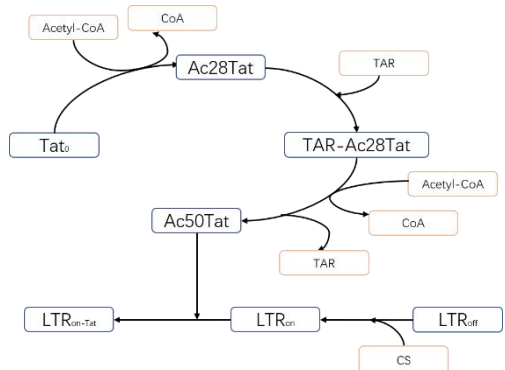

**Fig. 2 The flow chart of transcriptional-regulated chemical reactions on LTR**

## 1.2 常微分方程模型

控制和维持 HIV 功能的分子开关可简化地看作主要由 LTR 和宿主细胞因子、Tat 组成，LTR 的不同结

合状态如图 1 所示。未结合转录因子时的 LTR 状态记为 $LTR_{off}$；宿主细胞因子结合至 LTR 时促进转录起始，此状态记为 $LTR_{on}$；当宿主细胞因子和 Tat 均结合至 LTR 时促进转录起始和转录延伸，此状态记为 $LTR_{on\text{-}Tat}$。

病毒蛋白 Tat 促进转录延伸的正反馈机制中包含图 2 所示的多步修饰和结合反应，用变量 $x$ 表示处于不同状态 Tat 的总数量，在本模型中假设编码 Tat 的 RNA 数量与 TAR RNA 的数量相同，用变量 $y$ 表示。经由转录、翻译后生成的无修饰 Tat 记为 $Tat_0$，用 $a_0$ 表示其数量；K28 位点乙酰化的 Tat 记为 TatAc28，用 $a_1$ 表示其数量；处于与 TAR RNA 结合状态的 Tat 记为 TatAc28-TAR，用 $a_2$ 表示其数量；K50 位点乙酰化的 Tat 记为 TatAc50，用 $a_3$ 表示其数量；假设各状态 Tat 均可招募反应所需因子，用 $R$ 表示招募反应的平衡常数，则 $R \times x$ 为 Tat 所招募的因子如乙酰化辅酶（CoA）的数量。由 K28 位点乙酰化平衡常数 $k_1$、Tat 与 TAR RNA 结合平衡常数 $k_2$、K50 位点乙酰化平衡常数 $k_3$，可得平衡时各状态 Tat 与作用于 LTR 上促进转录延伸的 TatAc50 的关系为：

$$\begin{cases} a_0 = \frac{1}{k_1 k_2 k_3 R^2 x^2} \times a_3 \\ a_1 = \frac{1}{k_2 k_3 R x} \times a_3 \\ a_2 = \frac{y}{k_3 R x} \times a_3 \end{cases} \tag{1}$$

由 Tat 总数量守恒，即 $x = a_0 + a_1 + a_2 + a_3$，可得 TatAc50 与 Tat 总数量的关系为：

$$a_3 = \frac{k_1 k_2 k_3 R^2 \times x^3}{k_1 k_2 k_3 R^2 \times x^2 + k_1 k_2 R \times xy + k_1 R \times x + 1} \tag{2}$$

想象一个充满 LTR、宿主细胞因子和 Tat 的化学平衡系综；$C$ 表示宿主细胞转录因子数量的影响，即结合于 LTR 的宿主细胞转录因子（Sp1、NF-κB、NFAT 等）的数量；$l_1$、$l_2$、$l_3$ 分别表示三种不同结合状态即 $LTR_{off}$、$LTR_{on}$、$LTR_{on\text{-}Tat}$ 的频数。在某一个时间点，真实的情况是，3 种结合状态中有且仅有 1 种发生，在化学平衡的时候，各个结合状态的频数比值是恒定的，分别用 $P(1)$、$P(2)$、$P(3)$ 表示 LTR 处于 $LTR_{off}$、$LTR_{on}$、$LTR_{on\text{-}Tat}$ 结合状态的概率，此概率分布会随着蛋白质数量的改变而发生改变。

由宿主细胞因子结合于 LTR 的平衡常数 $K_1$ 和宿主细胞因子结合时 TatAc50 结合于 LTR 的平衡常数 $K_2$ 可得频数 $l_1$、$l_2$、$l_3$ 之间的关系；把三种结合状态的频数加起来作为一个配分函数，即归一化因子，则处于 $LTR_{off}$ 状态的概率 $P(1) = \frac{l_1}{Z}$、处于 $LTR_{on}$ 状态的概率 $P(2) = \frac{l_2}{Z}$、处于 $LTR_{on\text{-}Tat}$ 状态的概率 $P(3) = \frac{l_3}{Z}$。因此，LTR 处于每种状态的概率分布为：

$$\begin{cases} P(1) = \frac{1}{1 + K_1 \times C + K_1 K_2 \times C \times a_3} \\ P(2) = \frac{K_1 \times C}{1 + K_1 \times C + K_1 K_2 \times C \times a_3} \\ P(3) = \frac{K_1 K_2 \times C \times a_3}{1 + K_1 \times C + K_1 K_2 \times C \times a_3} \end{cases} \tag{3}$$

因此，Tat 数量和 RNA 数量变化速率分别为：

$$\begin{cases} f_x(x, y) = y \times L - \delta_t \times x \\ f_y(x, y) = N \times (T_1 \times P_1 + T_2 \times P_2 + T_3 \times P_3) - \delta_r \times y \end{cases} \tag{4}$$

其中，$L$ 表示翻译生成 Tat 的速率常数，$\delta_t$ 表示 Tat 降解的速率常数；$N$ 表示单个细胞中 HIV DNA 的数量，在本文的计算中将其取为 1[15]；$T_1$、$T_2$、$T_3$ 分别表示 LTR 处于 $LTR_{off}$、$LTR_{on}$、$LTR_{on\text{-}Tat}$ 3 种不同状态时的转录速率常数；$\delta_r$ 表示 RNA 降解的速率常数。由于 Tat 是由 HIV 编码的蛋白，因此可以其数量的高低作为病毒基因表达情况的表征，当 Tat 数量较高时，表示病毒处于转录活跃即快速复制的激活态，

反之则处于转录沉默的潜伏态。

## 1.3 动力学结构分解

随机性在生物学中是普遍存在的，本文模型里的随机性表现为细胞中大分子（蛋白质和 RNA）数量的涨落。在动力学中，分子数量涨落的量级为 $N^{1/2}$，并且其修正项的量级为 $1/N^{1/2}$（当 $N$ 特别大时可忽略），而在生物细胞中，大分子数目只有几十至几百个左右，因此其涨落是不可忽略的，即描述大分子数量随时间的变化的方程应包含随机性。对此，我们将使用以下随机微分方程作描述系统动力学：

$$\frac{dq}{dt} = f(q) + \zeta(q,t) \tag{5}$$

其中 $q^\tau = (x,y)$，$\tau$ 表示转置；$f^\tau(q) = (f_x, f_y)$ 表示改变大分子数量的确定性非线性因素；$\zeta(q,t)$ 表示系统噪声，假设其为均值为零的高斯白噪声，满足：

$$\langle \zeta(q,t)\zeta^\tau(q,t')\rangle = 2\varepsilon D(q)\delta(t-t') \tag{6}$$

其中正定对称矩阵 $D(q)$ 为扩散矩阵，$\varepsilon$ 为噪声强度。

对于以上随机微分方程存在唯一的动力学结构分解[16]：

$$[S(q) + A(q)]\frac{dq}{dt} = -\nabla U(q) + \xi(q,t) \tag{7}$$

其中正定对称矩阵 $S(q)$ 表示摩擦力，即耗散，对应于生物学中的降解，会促进系统趋于能量较低的稳态吸引子；反对称矩阵 $A(q)$ 表示横向力（洛伦兹力），对应生物学中的振荡，其不改变能量，对系统在不同稳态间的转换不起决定性作用，因此在本文所描述的开关现象中可将其忽略，但 $A(q)$ 在非点状吸引子系统中会对动力学行为产生影响，例如细胞周期的调控[17]；单值标量函数 $U(q)$ 表示势能函数；$\xi$ 表示随机力，与矩阵 $S(q)$ 存在随机耗散关系：

$$\langle \xi(q,t)\xi^\tau(q,t')\rangle = 2\varepsilon S(q)\delta(t-t') \tag{8}$$

将方程（5）代入方程（7），利用确定项的随机项的分别相等可以得到两个关系；随机项的相等引入了广义爱因斯坦关系：

$$[S(q) + A(q)]D(q)[S(q) - A(q)] = S(q) \tag{9}$$

确定项的相等得到：

$$[S(q) + A(q)]f(q) = -\nabla U(q) \tag{10}$$

为保证各部分的动力学性质的独立性，我们假设 $[S(q) + A(q)]$ 非奇异，即 $det[S(q) + A(q)] \neq 0$；因此方程（10）可变换为 $f(q) = -[S(q) + A(q)]^{-1}\nabla U(q) = -[D(q) + Q(q)]\nabla U(q)$，其中 $Q(q)$ 是由 $[D(q) + Q(q)] = [S(q) + A(q)]^{-1}$ 决定的反对称矩阵，在确定项的不动点处 $f(q) = 0$，由 $det[D(q) + Q(q)] \neq 0$ 可得势能函数 $U(q)$ 的极值点与确定性动力学不动点的对应与扩散矩阵 $D(q)$ 的具体形式无关，因此本文中为简化模型计算将其取为常数对稳态的分析不造成影响。

由势能函数的梯度的旋度为零的性质有：

$$\nabla \times [S(q) + A(q)]f(q) = 0 \tag{11}$$

方程（9）和（11）可由 $f(q)$ 和 $D(q)$ 求出 $S(q)$ 和 $A(q)$，从而进一步求得势能函数 $U(q)$：

$$U(q) = -\int [S(q) + A(q)]f(q)dq \tag{12}$$

由势能函数 $U(q)$ 可以确定分布函数：

$$\rho(q) = \rho_0 \exp\left[-\frac{U(q)}{\varepsilon}\right] \tag{13}$$

其中 $\rho_0$ 为归一化常数，$\rho_0 = \frac{1}{\int_0^{+\infty} \exp\left[-\frac{U(q)}{\varepsilon}\right]}$。使用这种随机积分方式求得的分布函数极值点与确定性部分常微分方程组的不动点一致，因此可以通过计算常微分方程组的稳定点分析系统的稳态情况。

## 2 计算结果

### 2.1 双稳态

本节计算中所代入的参数取值一部分通过查找相关文献获得，而另一部分未精确测量的参数取值，我们则根据文献中与其相关数据的大致数量级进行拟定，表1列出了模型中各参数的取值及相应参考文献。

**Table 1 Responses and parameters of positive feedback mechanism of HIV Tat**

| Reaction/Description | Parameters | Values | References |
|---|---|---|---|
| $Tat_0 + CoA \overset{k_1}{\rightleftharpoons} TatAc28$ | $k_1 = \dfrac{a_1}{a_0 \times R \times x}$ | $7.69 \times 10^{-3}$ | [10, 18] |
| $TatAc28 + TAR \overset{k_2}{\rightleftharpoons} TatAc28\text{-}TAR$ | $k_2 = \dfrac{a_2}{a_1 \times y}$ | $8.82 \times 10^{-3}$ | [10] |
| $TatAc28\text{-}TAR + CoA \overset{k_3}{\rightleftharpoons} TatAc50 + TAR$ | $k_3 = \dfrac{a_3 \times y}{a_2 \times R \times x}$ | $7.69 \times 10^{-3}$ | [10, 18] |
| $LTR_{off} + CS \overset{K_1}{\rightleftharpoons} LTR_{on}$ | $K_1 = \dfrac{l_2}{l_1 \times C}$ | $10^{-2}$ | [19] |
| $LTR_{on} + TatAc50 \overset{K_2}{\rightleftharpoons} LTR_{on\text{-}Tat}$ | $K_2 = \dfrac{l_3}{l_2 \times a_3}$ | $1$ | [19] |
| Transcription rate of $LTR_{off}$ | $T_1$ | $10^{-8} / s$ | [10, 18] |
| Transcription rate of $LTR_{on}$ | $T_2$ | $3.5 \times 10^{-6} / s$ | [19] |
| Transcription rate of $LTR_{on\text{-}Tat}$ | $T_3$ | $1.7 \times 10^{-4} / s$ | [19] |
| Translation rate of Tat | $L$ | $1.32 \times 10^{-3} / s$ | [18, 20] |
| Degradation rate of RNA | $\delta_r$ | $4.8 \times 10^{-5} / s$ | [10, 21] |
| Degradation rate of Tat | $\delta_t$ | $4.3 \times 10^{-5} / s$ | [10, 22] |
| Recruit equilibrium constant | $R$ | $15$ | [9] |
| Number of host transcription factors binding at LTR | $C$ | $85$ | [19] |

将表1中的参数取值代入方程（2）、（3）、（4）中，分别计算当 Tat 和 TAR RNA 初始值不同时其数量随时间的变化，得到结果如图3所示。同时，通过牛顿法可以计算得出 Tat 和 TAR RNA 在潜伏态不动点、激活态不动点和鞍点的数值，分别约为 1.0、78、35（Tat 数量）和 0.033、2.5、1.1（RNA 数量）。

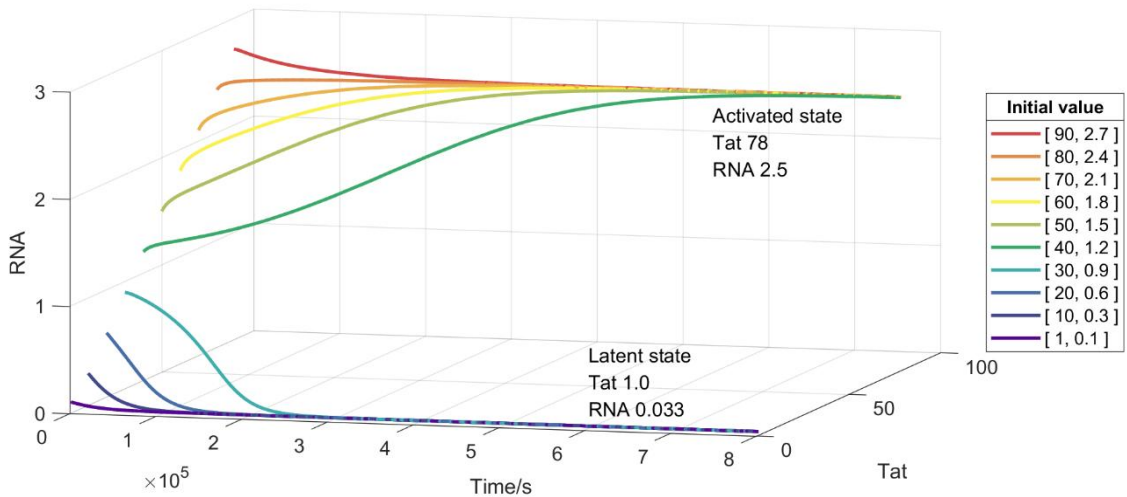

**Fig. 3 The number of Tat protein and RNA changed with time**

Note: The latent stable point was (1.0, 0.033), the active stable point was (78, 2.5) and the saddle point was (35, 1.1) by Newton's method.

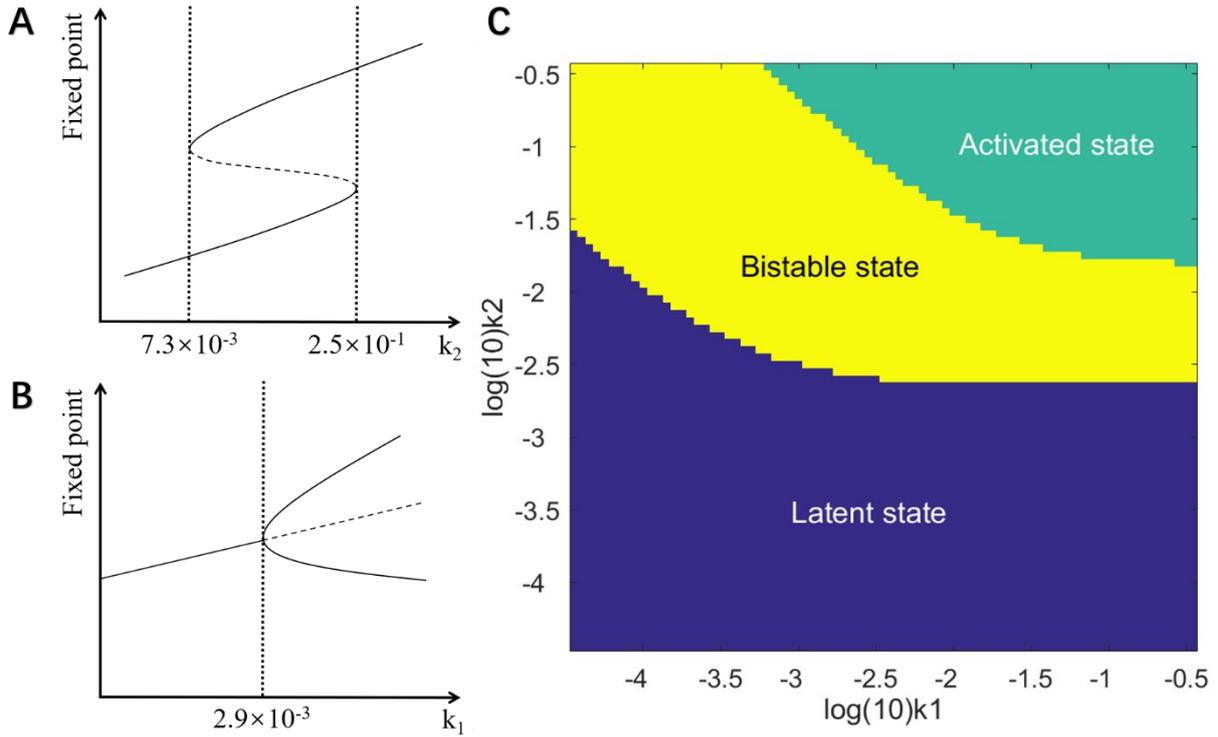

**Fig. 4 Bifurcation diagrams with parameters $k_1$ and $k_2$ as examples**

Note: (a, b) The solid line represents stable fixed point *i.e.* steady state, and the dotted line represents unstable fixed point *i.e.* saddle point. (c) Phase diagram with $k_1$ and $k_2$ as the parameter space, the blue area represents the monostable state with only latent state, the yellow area represents the bistable state with both latent state and active state, and the green area represents the monostable state with only activated state.

## 2.2 稳态个数变化

为了研究参数变化对稳态个数的影响，我们在保持其余参数不变的情况下，调整其中一个参数值，并通过作图得出在不同 Tat、RNA 初始值时 Tat 蛋白数量和 RNA 数量随时间的变化，同时用牛顿法计算出此时的稳态个数和相应数值（附件中第一部分），由此得到不同稳态个数所对应的参数区间。

根据参数的不同，系统可以表现出 3 种不同的稳态类型，以 $k_2$ 为例，其稳态分岔示意图见图 4a（对应 Tat 蛋白和 TAR RNA 结合的平衡常数对稳态个数的影响）：a. 只有潜伏态的单稳态，在 $k_2$ 较低（小于 $7.3\times10^{-3}$）时，只存在 1 个 Tat 和 RNA 数量较小的潜伏态；b. 只有激活态的单稳态，在 $k_2$ 较高（大于 $2.5\times10^{-1}$）时，只存在 1 个 Tat 和 RNA 数量较大的激活态；c. 双稳态，在上述两个 $k_2$ 值之间（$7.3\times10^{-3} \sim 2.5\times10^{-1}$）存在两个稳态，由初始 Tat 和 RNA 数量决定最终稳态：当初始数量较高时，最终 Tat 和 RNA 数量稳定在高表达数量的激活态，反之，则稳定在低表达数量的潜伏态。

我们继续调节其余参数，得到不同稳态个数的参数区间（取两位有效数字）如表 2 所示：

**表 2 不同稳态情况对应的参数区间**
**Table 2 Parameter ranges corresponding to different steady-state conditions**

| Parameters | Latent state | Bistable state | Activated state |
| --- | --- | --- | --- |
| $k_1$ | $<2.9\times10^{-3}$ | $\geq 2.9\times10^{-3}$ | $+\infty$ |
| $k_2$ | $<7.3\times10^{-3}$ | $7.3\times10^{-3} \sim 2.5\times10^{-1}$ | $>2.5\times10^{-1}$ |
| $k_3$ | $< 6.4 \times 10^{-3}$ | $6.4 \times 10^{-3} \sim 2.2 \times 10^{-1}$ | $> 2.2 \times 10^{-1}$ |
| $K_1$ | $< 7.4 \times 10^{-3}$ | $\geq 7.4 \times 10^{-3}$ | $+\infty$ |
| $K_2$ | $< 8.4 \times 10^{-1}$ | $8.4 \times 10^{-1} \sim 2.8 \times 10^{1}$ | $> 2.8 \times 10^{1}$ |
| $\delta_t$ | $> 4.7 \times 10^{-5}/s$ | $1.3 \times 10^{-5} \sim 4.7 \times 10^{-5}/s$ | $< 1.3 * 10^{-5}/s$ |
| $\delta_r$ | $> 5.2 \times 10^{-5}/s$ | $1.4 \times 10^{-5} \sim 5.2 \times 10^{-5}/s$ | $< 1.4 \times 10^{-5}/s$ |
| $C$ | $< 6.2 \times 10^{1}$ | $\geq 6.2 \times 10^{1}$ | $+\infty$ |
| $R$ | $< 1.3 \times 10^{1}$ | $1.3 \times 10^{1} \sim 1.2 \times 10^{2}$ | $> 1.2 \times 10^{2}$ |

由表 2 可知，当 $k_1$、$k_2$、$k_3$、$K_1$、$K_2$、$C$、$R$ 较低而 $\delta_t$、$\delta_r$ 较高时，只存在潜伏态；当 $k_2$、$k_3$、$K_2$、$R$ 较高而 $\delta_t$、$\delta_r$ 较低时，只存在激活态；另外，在本组参数下，单独调节 $k_1$、$K_1$、$C$ 时（已调至 $10^{20}$，在本文模型中将其看作无穷大），未出现只存在单激活态的情况（附件中第一部分），其中以 $k_1$ 为例的稳态分岔示意图如 4b 所示，以 $k_1$ 和 $k_2$ 为参数空间的相图如图 4c 所示。

在本小节中，我们通过调整不同的参数数值，分别得到了单潜伏态、双稳态和单激活态的情况，并且求出了不同稳态情况所对应的大致参数范围区间。

### 2.3 势能函数和概率分布

潜伏态和激活态之间的转换关系可以通过一个势能函数来表示，即 **1.3** 分解方法中的 $U(q)$。取扩散矩阵 $D(q)$ 为单位矩阵 $I$、噪声强度 $\varepsilon = 1$，使用该方法计算原始参数情况下潜伏态和激活态相对于鞍点的势能差分别为 $\Delta U_{b潜伏} = 0.0035$、$\Delta U_{b激活} = 0.0031$，即激活态比潜伏态更稳定。将鞍点处作为势能零点，相应的势能函数分布如图 5a 所示，其中的两个低谷分别对应于两个稳定状态，即潜伏态和激活态；通过 **1.3** 所述随机积分方法计算得到的 Tat 和 RNA 数目概率分布函数图像如图 5b 所示，其中的两个峰值分别对应于潜伏态和激活态。由于运用此种随机动力学结构分解的方法求得的概率分布函数极值点与确定性部分常微分方程组的不动点是一致的，说明加入噪声项不影响不动点的数值。图 6 为取 $\varepsilon = 5$ 时的势能函数与 Tat 和 RNA 数目的概率分布，对比图 5 和图 6 中稳态所对应的数值可以看出（由于只关注鞍点及稳定点附近的性质，图 5 中未展现出 $U > 5.0 \times 10^{-3}$，$P < 1.002 \times 10^{-4}$ 时其余地方的变化，图 6 中则未展现出 $U > 1.0 \times 10^{-3}$，$P < 1.002 \times 10^{-4}$ 时其余地方的变化），噪声强度的改变不影响不动点的数值。因此，仅需分析常微分方程确定性部分就可以求得稳态所对应 Tat 和 RNA 的数值。

另外，当参数改变时，相应的势能函数与 Tat 和 RNA 数目的概率分布会有所变化，可以从势能数值的高低和概率分布的大小来判断在不同参数下各稳态的稳定性，即当势能相对较低、分布概率高时，相应的稳态较稳定（附件中第二部分）。

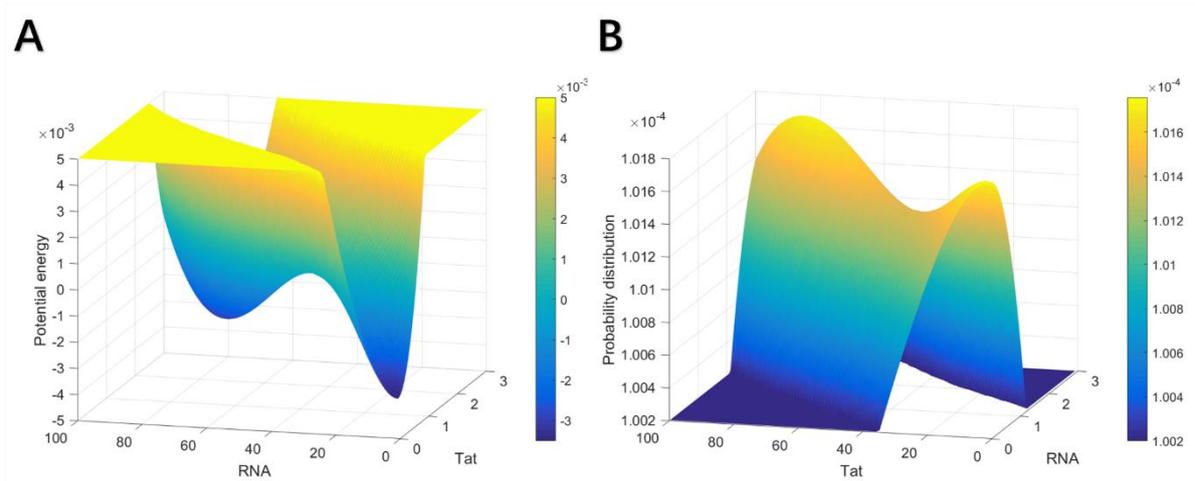

**Fig. 5** Potential energy function diagram when $\varepsilon = 1$ (a) and probability distribution diagram when $\varepsilon = 1$ (b)

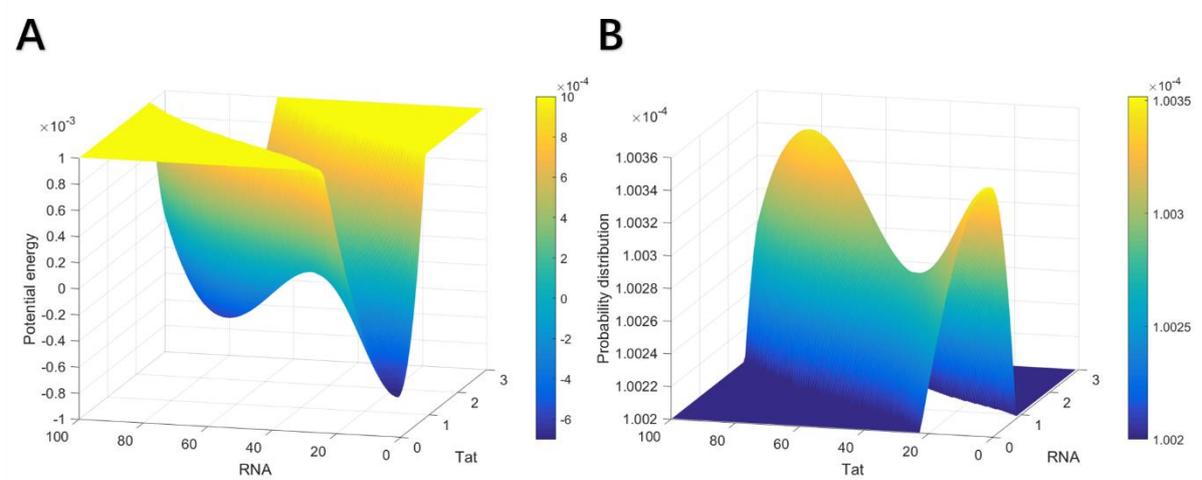

**Fig. 6** Potential energy function diagram when $\varepsilon = 5$ (a) and probability distribution diagram when $\varepsilon = 5$ (b), which are consistent with Fig. 5

## 3 讨 论

### 3.1 潜伏期建立机制

#### 3.1.1 稳态个数

系统的稳态个数发生变化表明其动力学结构出现分岔。本模型有仅存在潜伏态的单稳态、同时存在潜伏态和激活态的双稳态、仅存在激活态的单稳态 3 种不同的稳态情况：当系统仅存在潜伏态所对应的单稳态时，HIV 一定会进入潜伏期；当系统同时存在潜伏态和激活态的双稳态时，HIV 有一定几率进入潜伏期；当系统仅存在激活态所对应的单稳态时，HIV 一定不会进入潜伏期。由方程（10）可看出系统的稳态情况与随机性因素如扩散矩阵无关，仅与确定性因素如参数取值有关，各不同参数情况与稳态个数的对应如 **2.2** 中计算结果所示。

#### 3.1.2 稳定性分析

当系统同时存在潜伏态和激活态的双稳态时，不同稳态间的转换即为 HIV 在潜伏态与激活态之间的表型转换。根据 **2.3** 中的计算结果，不同参数情况下系统各稳态的势能不同，会导致处于各稳态的概率以及稳态间转换的概率不同。当潜伏态比激活态更稳定时，HIV 进入潜伏期的概率较大，并且更容易从激活态转换至潜伏态[23]；当激活态比潜伏态更稳定时，HIV 处于活跃复制的激活状态的概率较大，并且更容易从

潜伏态转换至激活态。系统各稳态的稳定性同时受到确定性因素和随机性因素的影响，但随机性的噪声强度仅影响势能的数值大小，并不改变极值点所对应的大分子数值，如图5、6所示，即HIV发生表型转换的概率会受到随机性因素的影响，但处于稳定状态时的大分子数量与随机性因素无关。

### 3.2 影响因素

#### 3.2.1 宿主细胞状态

根据 1.2 的模型，方程中的参数 $C$ 和 $K_1$ 分别表示宿主细胞因子的数量以及宿主细胞因子对 LTR 的作用程度，表征了宿主细胞状态在转录过程中的影响。由 2.2 中的计算结果，可以得出的结论是：当宿主细胞因子数量较少、对 LTR 作用程度较弱，即宿主细胞处于静息状态时，仅会存在潜伏态的单稳态；当宿主细胞因子数量增加、对 LTR 作用程度增强，即宿主细胞状态转变为活跃状态时，系统会出现潜伏态和激活态并存的双稳态。在双稳态区间内，宿主细胞状态的改变不会影响病毒表达，即病毒的状态可以独立于细胞状态自主调控，与实验现象[19]一致；而当宿主细胞状态的变化超出了双稳态区间时，就会对病毒的表达情况产生影响。但在本文的参数条件下，$C$ 和 $K_1$ 的增加均不会导致仅存在激活态的单稳态出现，即无法使潜伏态的稳态消失，说明该状况下宿主细胞的激活无法彻底清除病毒存储库。

#### 3.2.2 Tat 正反馈机制

在 Tat 正反馈的反应过程中，Tat 的乙酰化难易程度、Tat 与 TAR RNA 的亲和力、Tat 降解速率与招募反应效果均会影响正反馈的强弱。

在 1.2 的模型中，当 $k_2$、$k_3$ 较小，即 Tat-TAR 结合速率较低而分离速率较高、TatK50位点的去乙酰化速率较高而乙酰化速率较低时，Tat 正反馈强度较小，系统仅存在潜伏态的单稳态，随着 $k_2$、$k_3$ 的增大，系统会出现潜伏态和激活态并存的双稳态，继续增大则会仅剩下激活态的单稳态；招募反应平衡常数 $R$ 的增大也会导致系统的动力学结构出现同样的分岔行为。但由 2.2 中的计算结果可以看出，Tat 在 K28 位点的乙酰化对系统稳态个数的影响程度较弱，平衡常数 $k_1$ 的增加不会使潜伏态的稳态消失。

而当 Tat 的降解速率 $\delta_t$ 较小，即 Tat 半衰期较长时，Tat 正反馈强度较大，系统仅存在激活态的单稳态，随着 $\delta_t$ 的增大，系统会出现潜伏态和激活态并存的双稳态，继续增大则会仅剩下潜伏态的单稳态。因此，当 Tat 半衰期的变化较剧烈，越过了分岔临界值时，HIV 表现出的稳态类型就会发生变化，即改变 Tat 正反馈的强度可以在不激活细胞状态的情况下使病毒表达在潜伏态与激活态之间转换，与实验现象[19]一致。

#### 3.2.3 噪声

将 2.1 中由常微分方程计算得到的稳定点数值与 2.3 中分解随机微分方程计算出的势能和概率分布的局部极值点对比，可以验证稳态所对应的大分子数值仅由确定性因素决定，即噪声不影响系统动力学结构的分岔，因此当系统仅存在单稳态时无论噪声强度如何 HIV 都不会发生表型转换；但噪声会影响势能和概率分布的数值大小，即处于双稳态区间的系统发生表型转换的难易程度同时受到确定性因素和随机性因素的影响，而当确定性因素不变时，能否在足够短的时间内发生表型转换受到噪声的驱动，由于生物系统中噪声的普遍存在，实验上可以观察到 HIV 在潜伏态与激活态之间随机转换[18]。需要强调的是，噪声只是导致表型转换而并非产生双稳态的原因，双稳态的存在由确定性动力学的分岔结构决定。

## 4 结 论

HIV 潜伏库的存在极大阻碍了艾滋病的治愈，本研究通过建立调控 HIV 转录过程的分子开关的动力学模型定量分析了潜伏期的建立机制，并探究了不同因素对系统稳态情况的影响，对激活病毒潜伏库的临床治疗具有指导意义。我们模型中所使用的连续的随机微分方程相较于以往离散的方法在很大程度上增加了变量和参数的探索空间，有利于进一步扩大模型范围，并在一定程度上促进了寻找彻底治愈艾滋病的有效

的治疗靶点与药物的进程。

利用本文中的模型，可以计算出在不同参数下，系统动力学结构所存在的稳态情况，即处于单稳态或者双稳态，在此基础上，可得到各稳态所对应的势能，并通过不同参数对稳态数量以及势能大小的影响评估不同药物靶点的治疗程度，例如当调节参数取值越过分岔临界值，使系统由双稳态变为仅存在激活态的单稳态时，该参数所对应的药物治疗即可达到彻底激活病毒库的效果。对比 **2.2** 计算结果中各参数在双稳态区间取值范围，我们可以初步得出，宿主细胞状态和 Tat 在 K28 位点乙酰化的难易程度对稳态个数变化的影响程度较弱。因此，在本文所拟定的参数调节下，应更着重于发展调控 Tat-TAR 亲和力、TatK50 位点乙酰化、Tat 的招募反应效果以及 Tat 半衰期的药物靶点，与目前已有的生物学实验结果一致，验证了本工作的理论基础。当临床上可以测得不同病人的体内参数具体数值时，则可通过本文的方法计算分析稳态及势能的情况，从而在理论上得出更加有效的药物靶点，指导制定针对相应病人的个性化治疗方案。

总的来说，我们的研究结果表明，在建立调控 HIV 转录的分子开关的随机动力学模型的基础上，通过定量分析方程中确定性部分的动力学结构，可以得到潜伏期建立的分子机制及各药物靶点的治疗效果，有利于进一步了解艾滋病在生物体内的发生发展机制，为临床治疗提供理论指导。

**附件** 详细计算结果

# A Quantitative Analysis of Dynamic Mechanisms Regulating HIV Latency and Activation[*]

XIONG Rui-Qi[**], SU Yang[**], AO Ping[***]

(*Shanghai Center for Quantitative Life Sciences and Physics Department*, *Shanghai University*, *Shanghai* 200444, China)

**Abstract** **Objective** The reservoir of human immunodeficiency virus (HIV) latently infected cells is the major obstacle for eradication of acquired immunodeficiency syndrome (AIDS). Due to the noisy environment and multiple influencing factors in the organism, current dynamical models cannot reach a common understanding of the molecular mechanism of HIV latency. In this work, through a new dynamical structure decomposition, the deterministic part of the equation can be separated from the stochastic noise. Thus, the fixed-point analysis of ordinary differential equation is enough to obtain the different steady states of the system. **Methods** We established a dynamical model of HIV transcription process by using continuous stochastic differential equations, which simplifies the dimensions of equations needed to describe the system and increases the explorable space of the model. Different states between latency and activation of virus and their relations can be intuitively represented by potential functions and probability distribution functions. **Results** Based on our model, the influence of different dynamical parameters on stability is quantitatively analyzed, the parameter ranges of the system in bistable and monostable states are obtained respectively. The theoretical basis of this work is verified by comparing the effects of different factors on dynamic bifurcation with the results of biological experiments. **Conclusion** This paper goes beyond previous discrete stochastic methods, and can quantitatively analyze the dynamic mechanism of HIV transcriptional regulation through ordinary differential equations, which is beneficial to the promotion to deal with the high-dimensional situation, and further study the occurrence and development of AIDS *in vivo*, so as to guide the design of experiments and search for clinical treatment.

**Key words** human immunodeficiency virus (HIV), ordinary differential equation (ODE), dynamical structure decomposition, bistability, molecular switch, potential function

[*] This work was supported in part by the National Natural Science Foundation of China No. 16Z103060007.
[**] These authors contributed equally to this work.
[***] Corresponding author.
Tel: 021-66137278, E-mail: aoping@sjtu.edu.cn

**Graphical abstract**

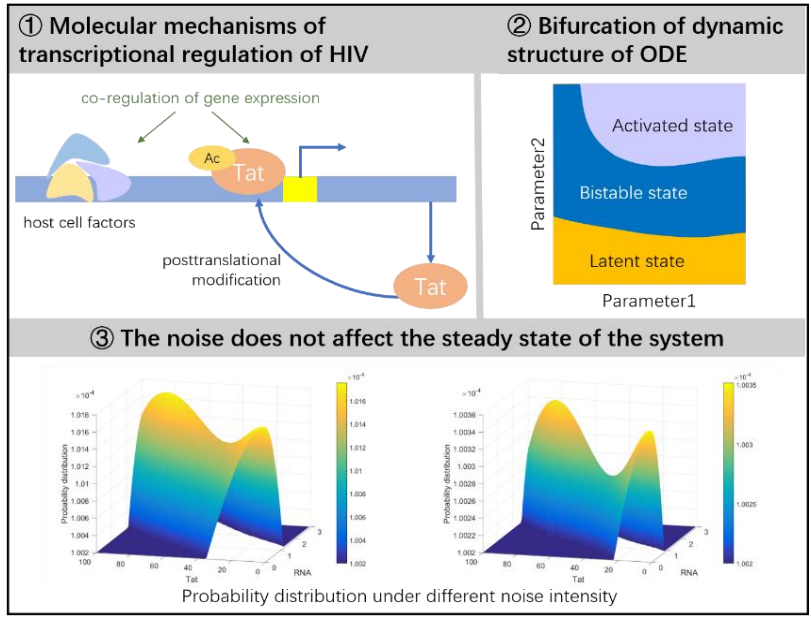

**Highlights**

・HIV expression status is related to host cell status and Tat positive feedback

・The bifurcation of the dynamical structure of the system can be analyzed by calculating the deterministic part of the stochastic differential equation

・Noise does not affect the steady-state solution of the system

**In Brief**

This work breaks through the previous discrete and stochastic methods, and uses continuous ordinary differential equations to describe the molecular regulation mechanism of HIV steady-state transition. Different expression states between latency and activation of virus and their relations can be quantitatively analyzed by potential functions and probability distribution functions.

# 附件 详细计算结果

## 第一部分 不同参数条件下 Tat 蛋白数量随时间变化情况

在保持正文中其余参数不变的情况下，调整其中一个参数值，通过作图法记录此时的稳态个数，并用牛顿法计算此时稳态对应的 Tat 和 TAR RNA 的数值。

a. 调整参数 $k_1$，出现只有单潜伏态的情形（在此组参数下，无法得出单激活态的情形）

Ⅰ：取 $k_1 = 10^{-3}$ 时，只有潜伏态（图 S1）

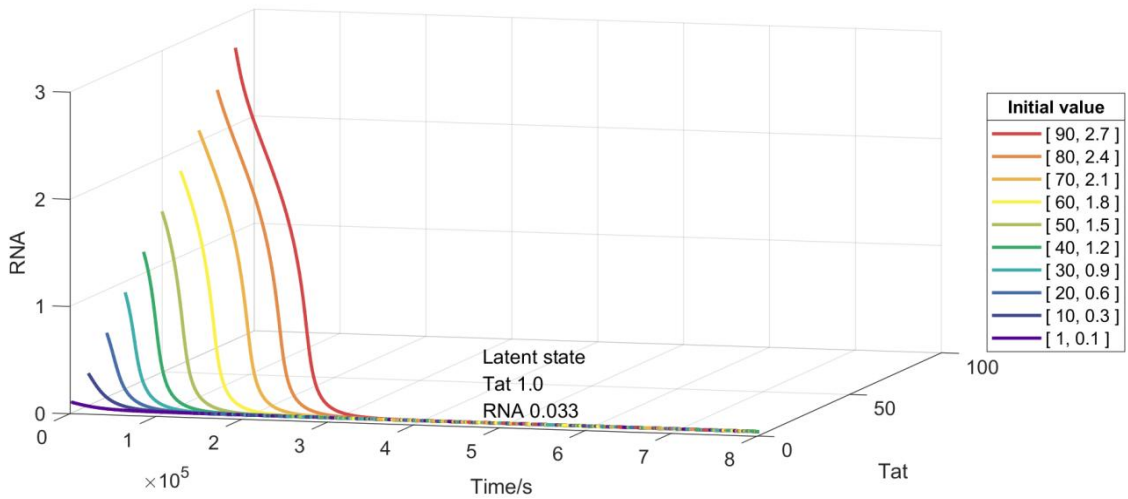

**Fig. S1** The number of Tat protein and RNA changed with time, and was (1.0, 0.033) at latent stable point

潜伏态稳定点 Tat 数值为 1.0，RNA 数值为 0.033

b. 调整参数 $k_2$，出现只有单激活态和单潜伏态的情形

Ⅰ：取 $k_2 = 1$ 时，只有激活态（图 S2）

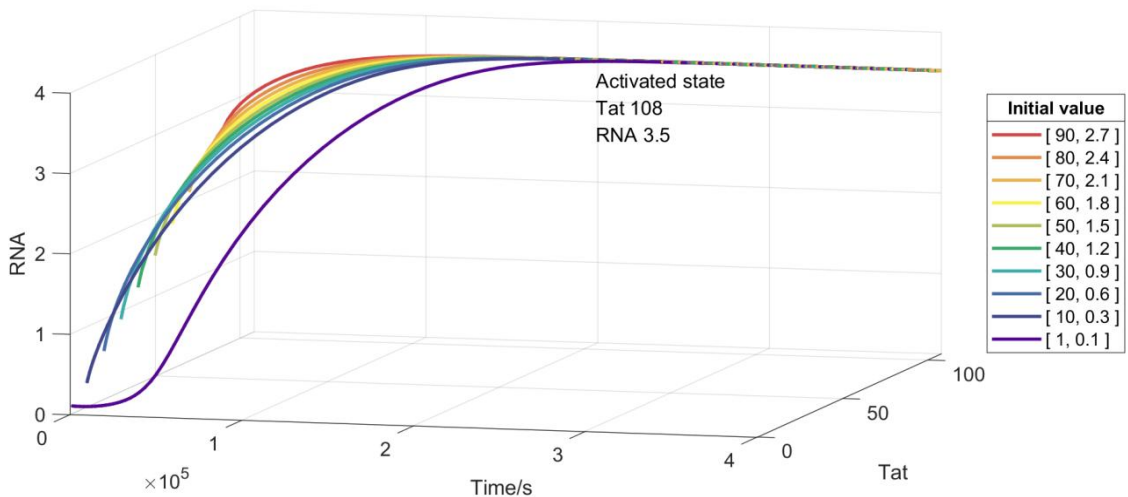

**Fig. S2** The number of Tat protein and RNA changed with time, and was (108, 3.5) at active stable point

激活态稳定点 Tat 数值为 108，RNA 数值为 3.5

Ⅱ：取 $k_2 = 10^{-3}$ 时，只有潜伏态（图 S3）

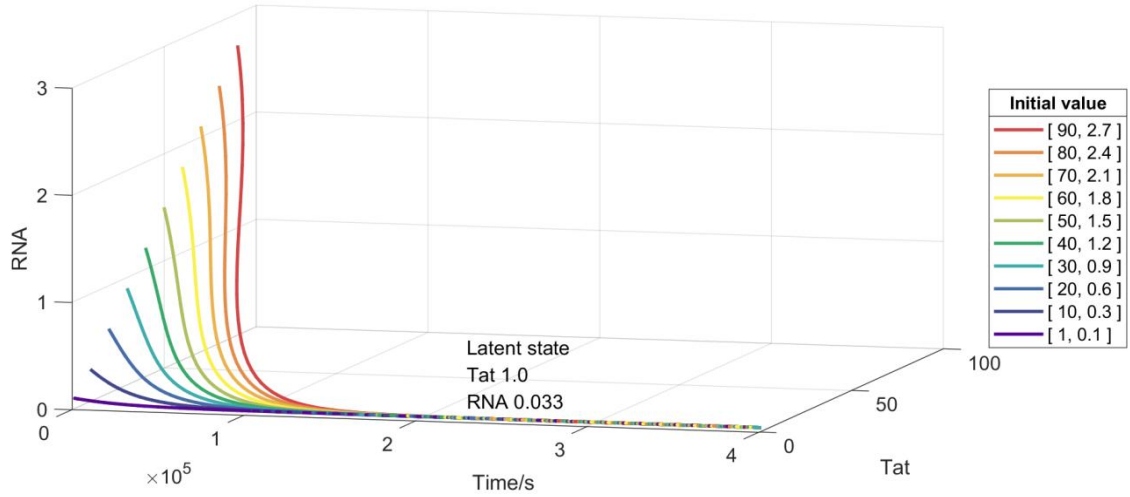

**Fig. S3  The number of Tat protein and RNA changed with time, and was (1.0, 0.033) at latent stable point**

**潜伏态稳定点 Tat 数值为 1.0，RNA 数值为 0.033**

c．调整参数 $k_3$，出现只有单激活态和单潜伏态的情形

　Ⅰ：取 $k_3 = 1$ 时，只有激活态（图 S4）

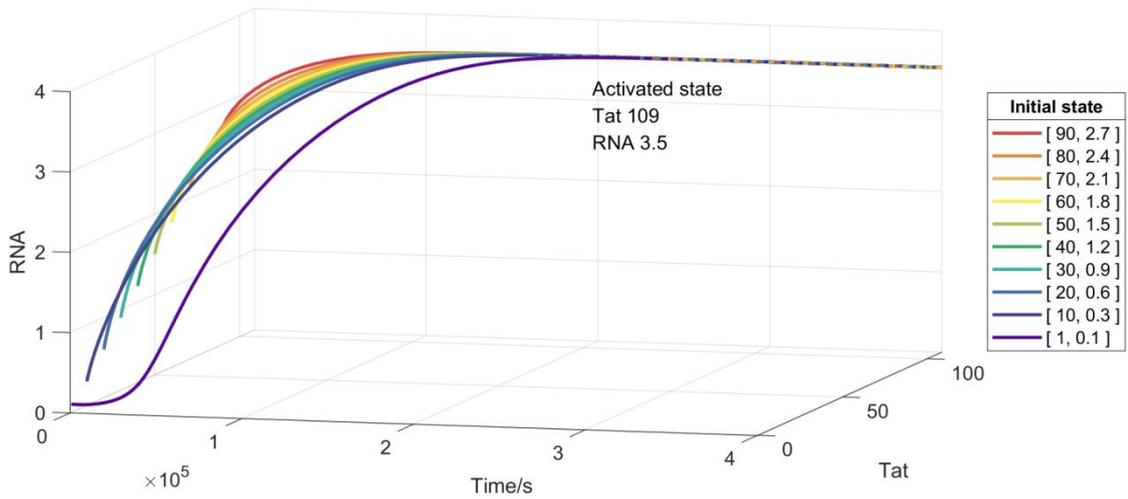

**Fig. S4  The number of Tat protein and RNA changed with time, and was (109, 3.5) at active stable point**

**激活态稳定点 Tat 数值为 109，RNA 数值为 3.5**

　Ⅱ：取 $k_3 = 10^{-3}$ 时，只有潜伏态（图 S5）

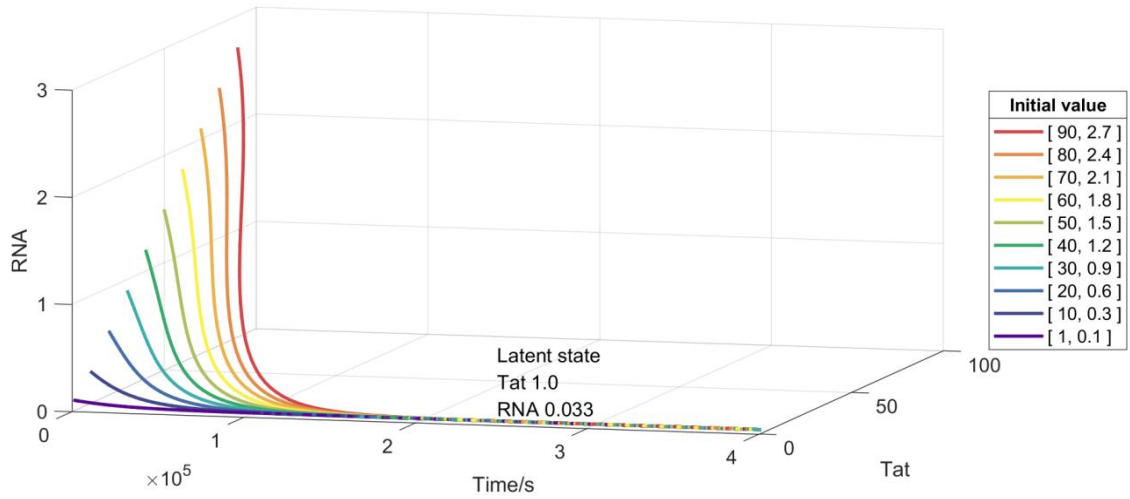

**Fig. S5  The number of Tat protein and RNA changed with time, and was (1.0, 0.033) at latent stable point**

潜伏态稳定点 Tat 数值为 1.0，RNA 数值为 0.033

d. 调整参数 $K_1$，出现只有单潜伏态的情形（在此组参数下，无法得出单激活态的情形）

　　Ⅰ：取 $K_1 = 10^{-3}$ 时，只有潜伏态（图 S6）

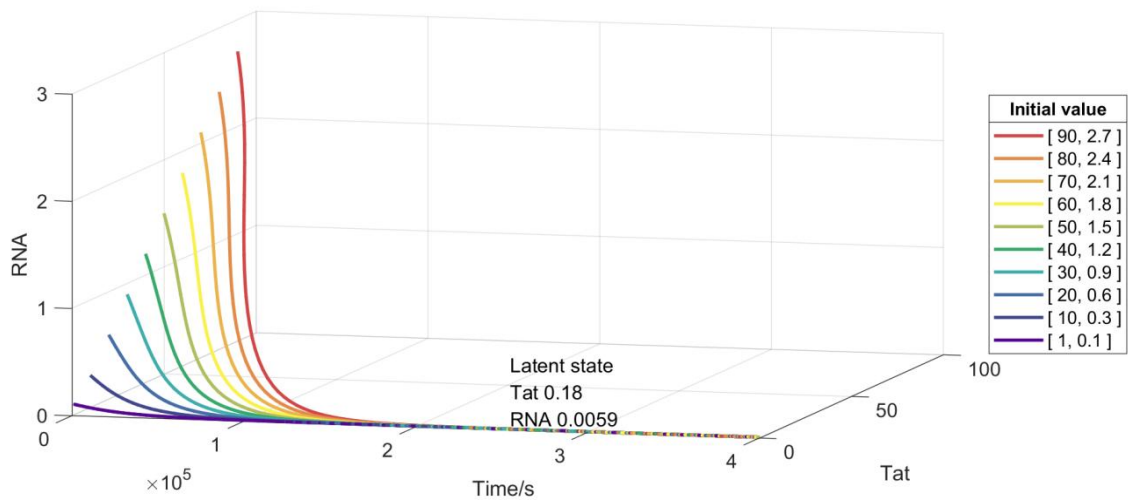

**Fig. S6  The number of Tat protein and RNA changed with time, and was (0.18, 0.0059) at latent stable point**

潜伏态稳定点 Tat 数值为 0.18，RNA 数值为 0.0059

e. 调整参数 $K_2$，出现只有单激活态和单潜伏态的情形

　　Ⅰ：取 $K_2 = 40$ 时，只有激活态（图 S7）

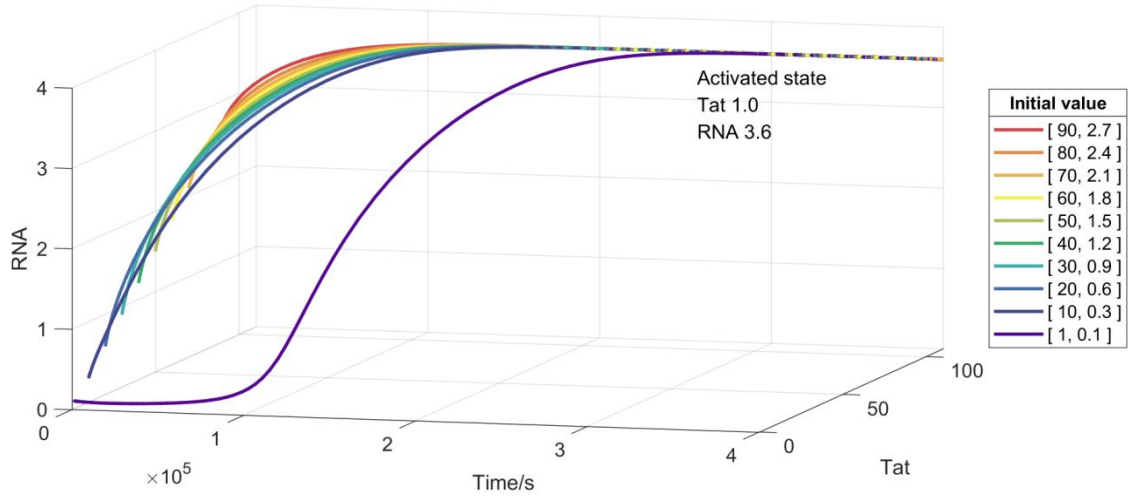

**Fig. S7　The number of Tat protein and RNA changed with time, and was (110, 3.6) at active stable point**
激活态稳定点 Tat 数值为 110，RNA 数值为 3.6

Ⅱ：取 $K_2 = 10^{-3}$ 时，只有潜伏态（图 S8）

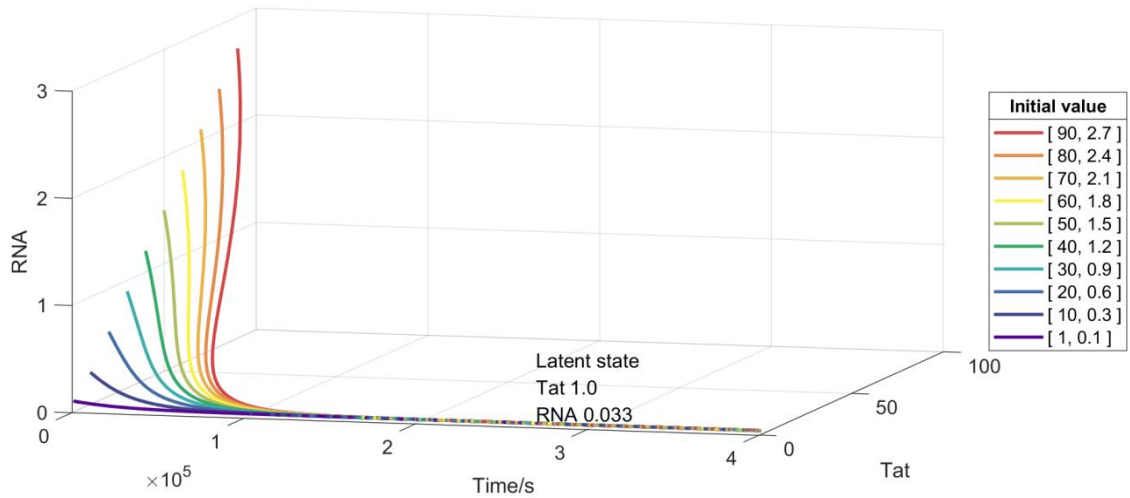

**Fig. S8　The number of Tat protein and RNA changed with time, and was (1.0, 0.033) at latent stable point**
潜伏态稳定点 Tat 数值为 1.0，RNA 数值为 0.033

f．调整参数 $C$，出现只有单潜伏态的情形（在此组参数下无法得到单激活态的情形）

　　Ⅰ：取 $C = 50$ 个 时，只有潜伏态（图 S9）

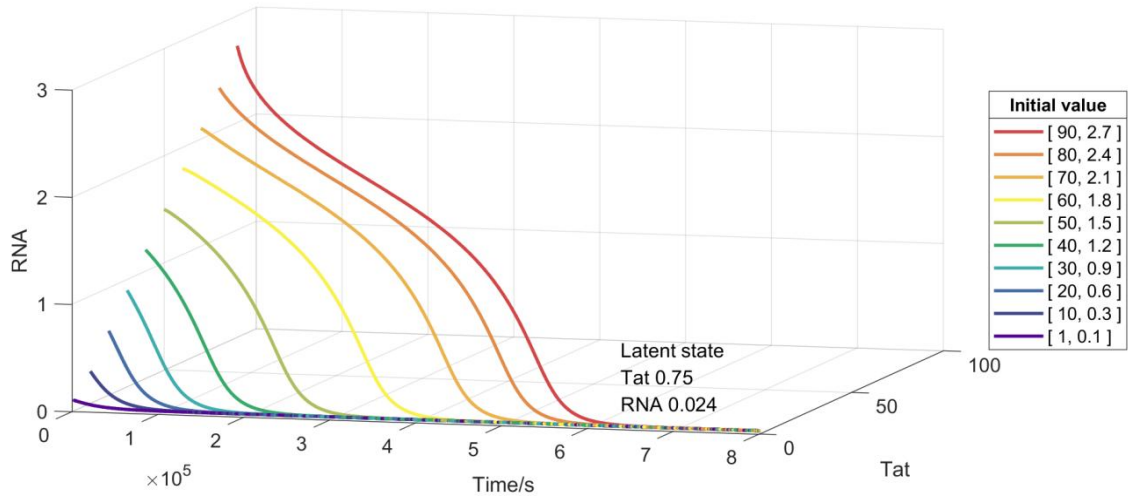

**Fig. S9    The number of Tat protein and RNA changed with time, and was (0.75, 0.024) at latent stable point**

潜伏态稳定点 **Tat** 数值为 **0.75**，**RNA** 为 **0.024**

g. 调整参数 $r$，出现只有单激活态和单潜伏态的情形

Ⅰ：取 $r = 150/s$ 时，只有激活态（图 S10）

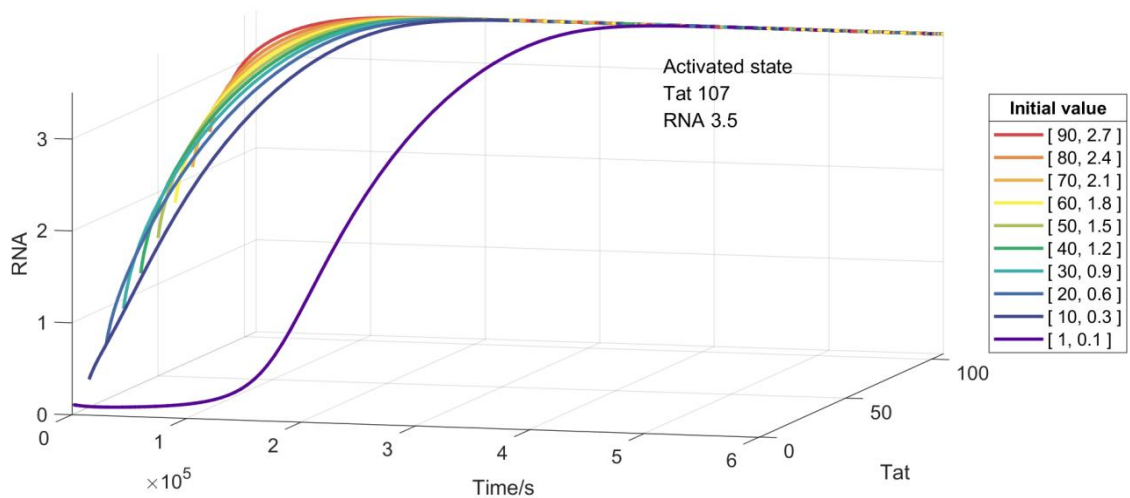

**Fig. S10    The number of Tat protein and RNA changed with time, and was (107, 3.5) at active stable point**

激活态稳定点 **Tat** 数值为 **107**，**RNA** 等于 **3.5**

Ⅱ：取 $r = 10/s$ 时，只有潜伏态（图 S11）

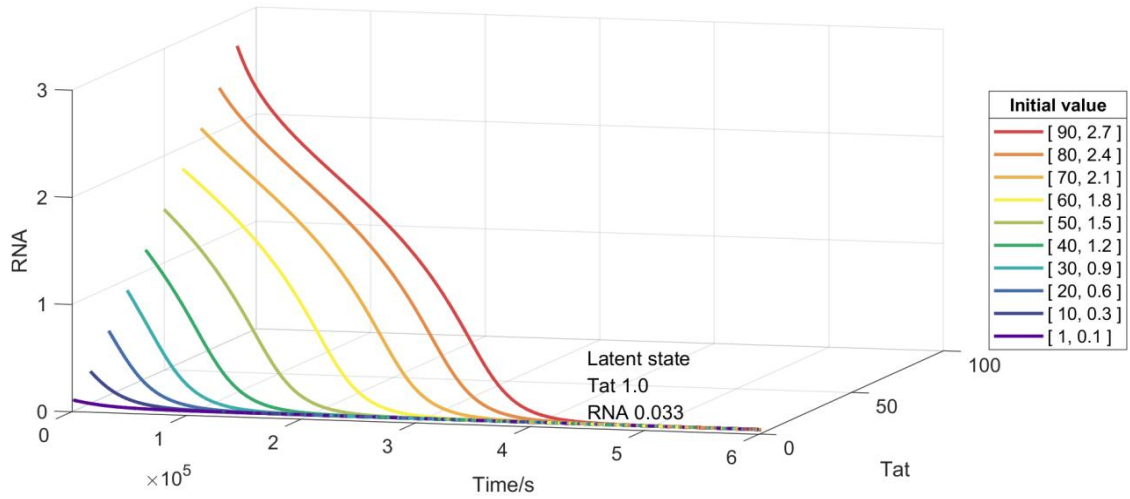

**Fig. S11　The number of Tat protein and RNA changed with time, and was (1.0, 0.033) at latent stable point**
潜伏态稳定点 Tat 数值为 1.0，RNA 等于 0.033

h. 调整参数 $\delta_t$，出现只有单激活态和单潜伏态的情形

　　Ⅰ：取 $\delta_t = 10^{-5}/s$ 时，只有激活态（图 S12）

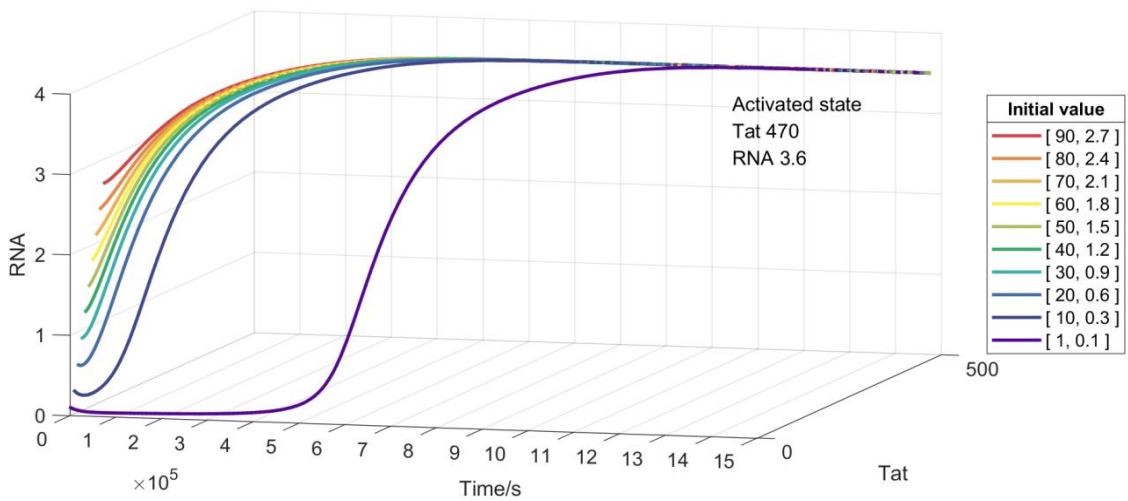

**Fig. S12　The number of Tat protein and RNA changed with time, and was (470, 3.6) at active stable point**
激活态稳定点 Tat 数值为 470，RNA 数值为 3.6

　　Ⅱ：取 $\delta_t = 6*10^{-5}$ 时，只有潜伏态（图 S13）

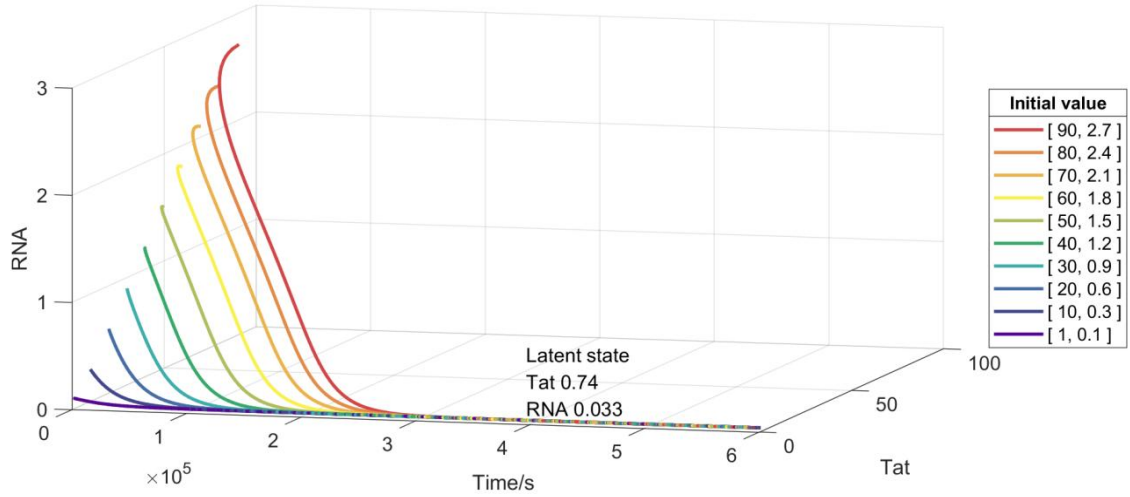

**Fig. S13    The number of Tat protein and RNA changed with time, and was (0.74, 0.033) at latent stable point**

潜伏态稳定点 Tat 数值为 0.74，RNA0.033

i. 调整参数 $\delta_r$，出现只有单激活态和单潜伏态的情形

Ⅰ：取 $\delta_r = 10^{-5}/s$ 时，只有激活态（图 S14）

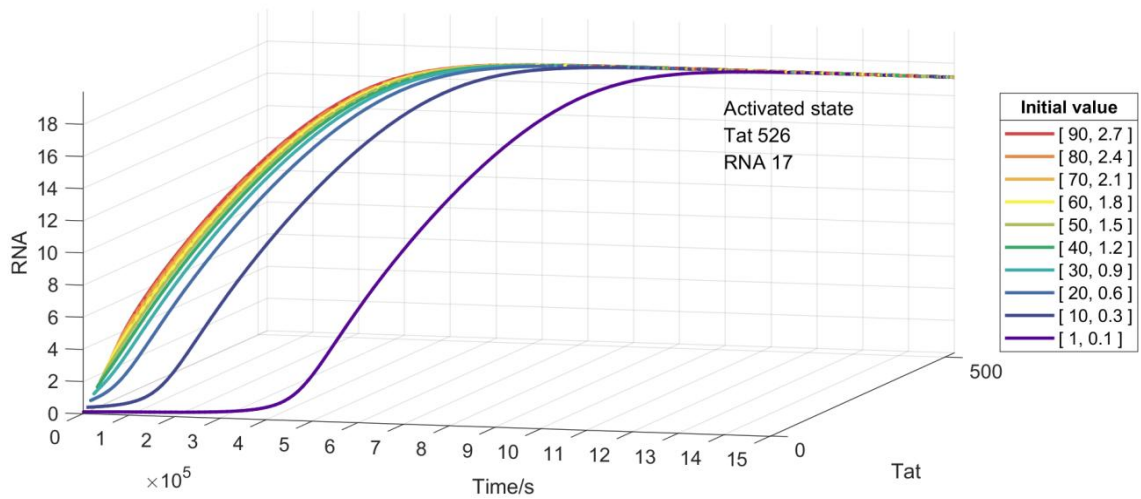

**Fig. S14    The number of Tat protein and RNA changed with time, and was (526, 17) at active stable point**

激活态稳定点 Tat 数值为 526，RNA 数值为 17

Ⅱ：取 $\delta_r = 6*10^{-5}$ 时，只有潜伏态（图 S15）

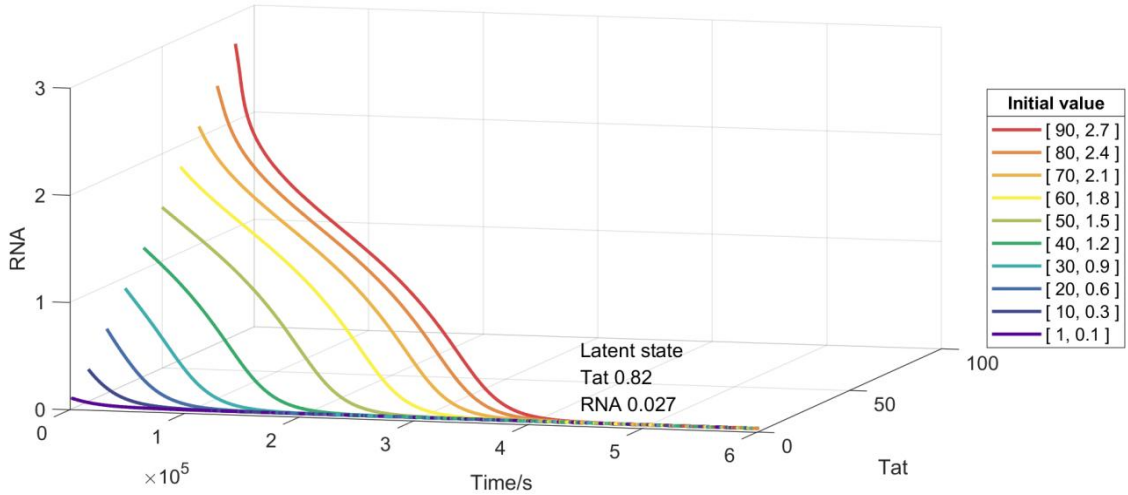

**Fig. S15** The number of Tat protein and RNA changed with time, and was (0.82, 0.027) at active stable point
潜伏态稳定点 Tat 数值为 0.82，RNA 数值为 0.027

## 第二部分 不同参数条件下的势能分布和 Tat 蛋白数量分布情况

在保持正文中其余参数不变的情况下，调整其中一个参数值，通过作图得出相应的稳态势能分布与 Tat 蛋白和 TAR RNA 数目的分布，从势能的高低（势能越低越稳定）和数目分布的大小（分布越多越稳定）来判断在此组参数下相应稳态的稳定性。

a．调整参数 $k_1$，出现只有单潜伏态的情形（在此组参数下，无法得出单激活态的情形）（取 Tat=50，RNA=1.5 时的势能为零点，并且由于只关注稳定点附近的性质，图中未展现出 $U > 5.0 \times 10^{-3}$，$P < 7 \times 10^{-5}$ 时其余地方的变化）

Ⅰ：取 $k_1 = 10^{-3}$ 时，只有潜伏态（图 S16）

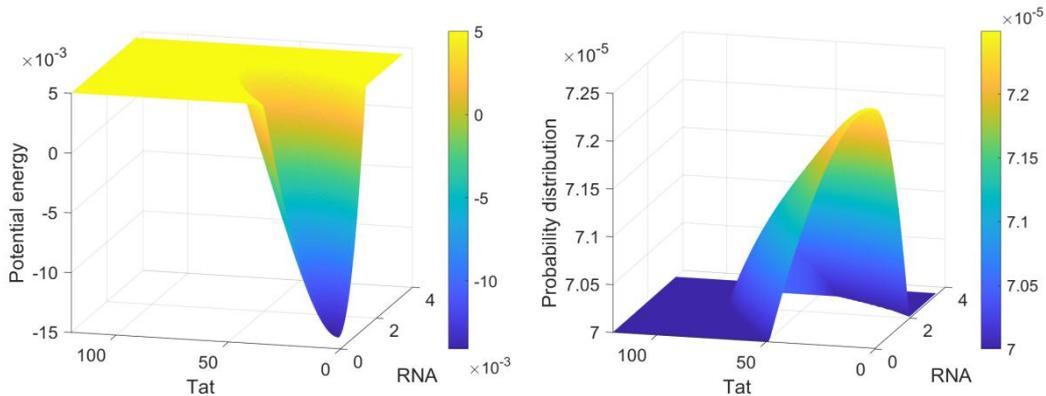

**Fig. S16** Potential energy function and probability distribution diagram when $k_1 = 10^{-3}$

b．调整参数 $k_2$，出现只有单激活态和单潜伏态的情形（取 Tat=50，RNA=1.5 时的势能为零点，并且由于只关注稳定点附近的性质，图中未展现出 $U > 5.0 \times 10^{-3}$，$P < 7 \times 10^{-5}$ 时其余地方的变化）

Ⅰ：取 $k_2 = 1$ 时，只有激活态（图 S17）

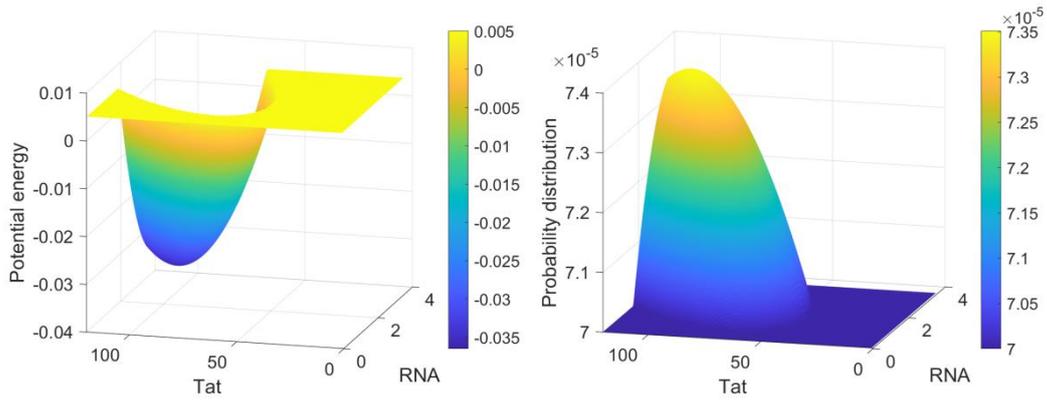

**Fig. S17** Potential energy function and probability distribution diagram when $k_2 = 1$

Ⅱ：取 $k_2 = 10^{-3}$ 时，只有潜伏态（图 S18）

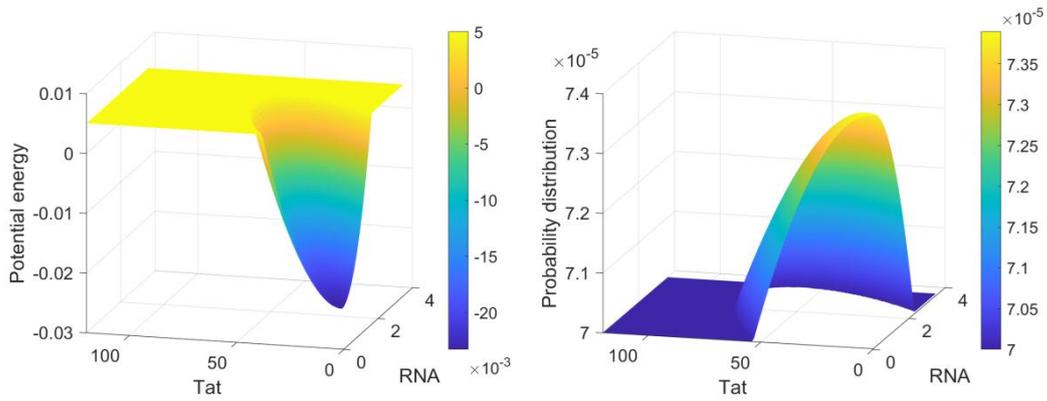

**Fig. S18** Potential energy function and probability distribution diagram when $k_2 = 10^{-3}$

c．调整参数 $k_3$，出现只有单激活态和单潜伏态的情形（取 Tat=50，RNA=1.5 时的势能为零点，并且由于只关注稳定点附近的性质，图中未展现出 $U > 5.0 \times 10^{-3}$，$P < 7 \times 10^{-5}$ 时其余地方的变化）

Ⅰ：取 $k_3 = 1$ 时，只有激活态（图 S19）

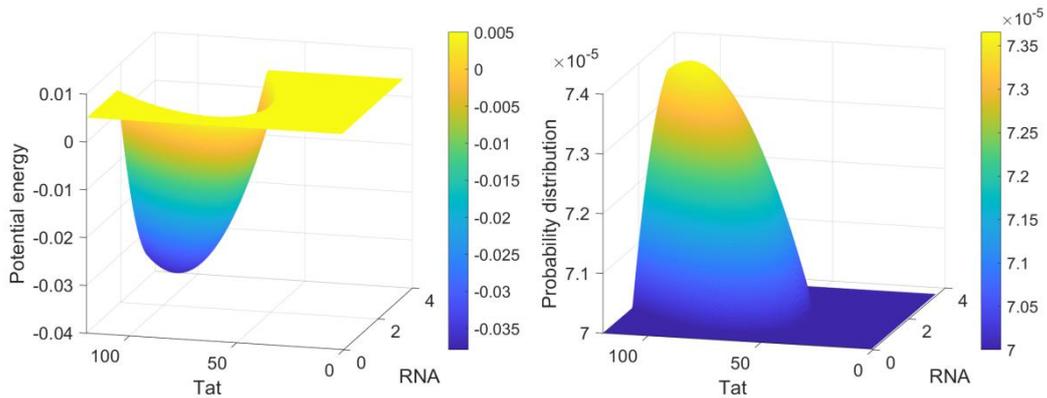

**Fig. S19** Potential energy function and probability distribution diagram when $k_3 = 1$

Ⅱ：取 $k_3 = 10^{-3}$ 时，只有潜伏态（图 S20）

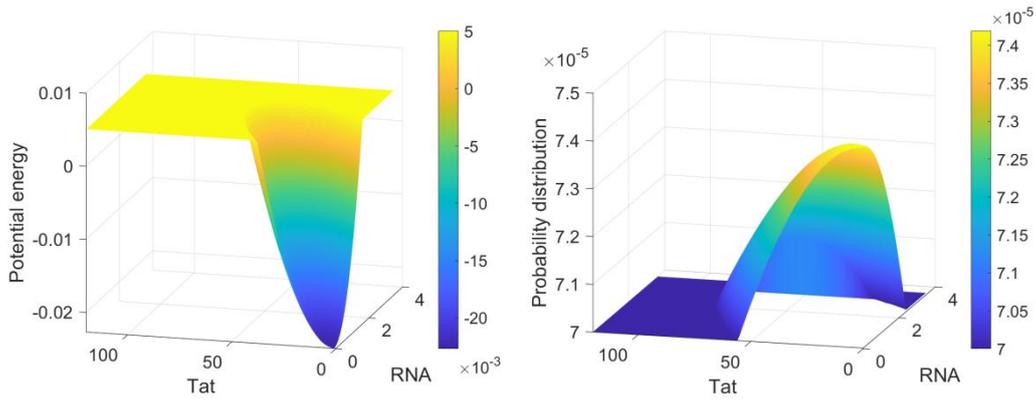

**Fig. S20    Potential energy function and probability distribution diagram when $k_3 = 10^{-3}$**

d．调整参数 $K_1$，出现只有单潜伏态的情形（在此组参数下，无法得出单激活态的情形）（取 Tat=50，RNA=1.5 时的势能为零点，并且由于只关注稳定点附近的性质，图中未展现出 $U > 5.0 \times 10^{-3}$，$P < 7 \times 10^{-5}$ 时其余地方的变化）

Ⅰ：取 $K_1 = 10^{-3}$ 时，只有潜伏态（图 S21）

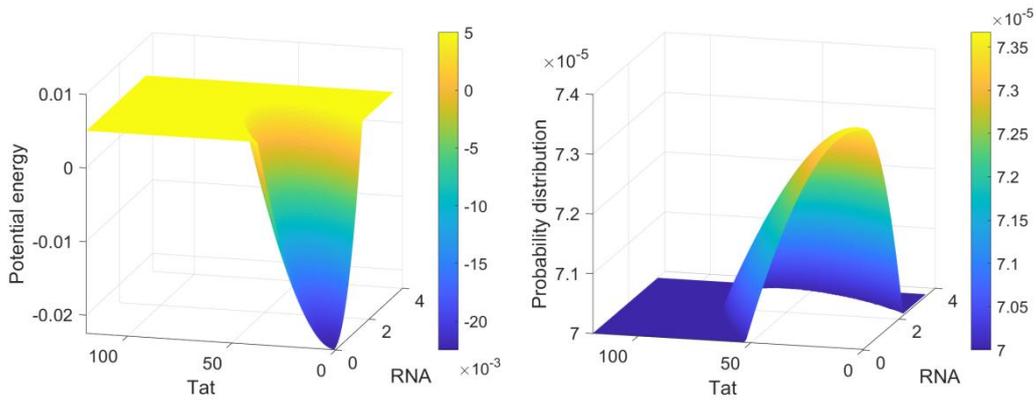

**Fig. S21    Potential energy function and probability distribution diagram when $K_1 = 10^{-3}$**

e．调整参数 $K_2$，出现只有单激活态和单潜伏态的情形（取 Tat=50，RNA=1.5 时的势能为零点，并且由于只关注稳定点附近的性质，图中未展现出 $U > 5.0 \times 10^{-3}$，$P < 7 \times 10^{-5}$ 时其余地方的变化）

Ⅰ：取 $K_2 = 40$ 时，只有激活态（图 S22）

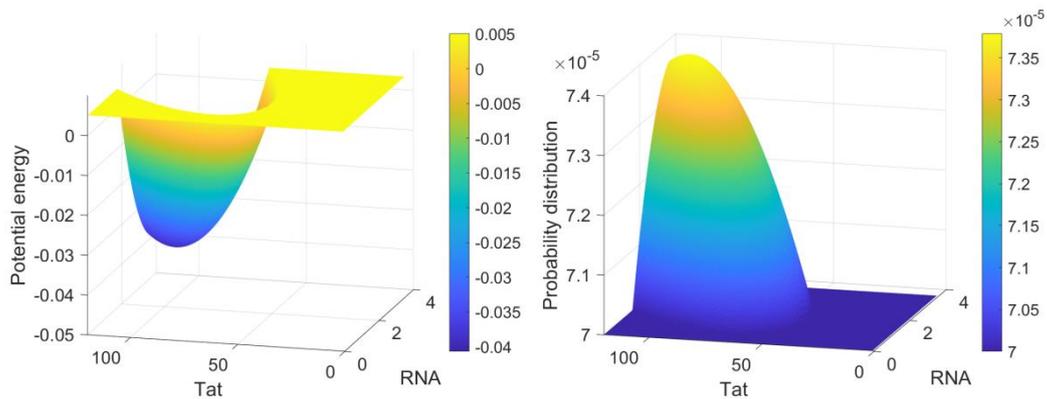

**Fig. S22    Potential energy function and probability distribution diagram when $K_2 = 40$**

Ⅱ：取 $K_2 = 10^{-3}$ 时，只有潜伏态（图 S23）

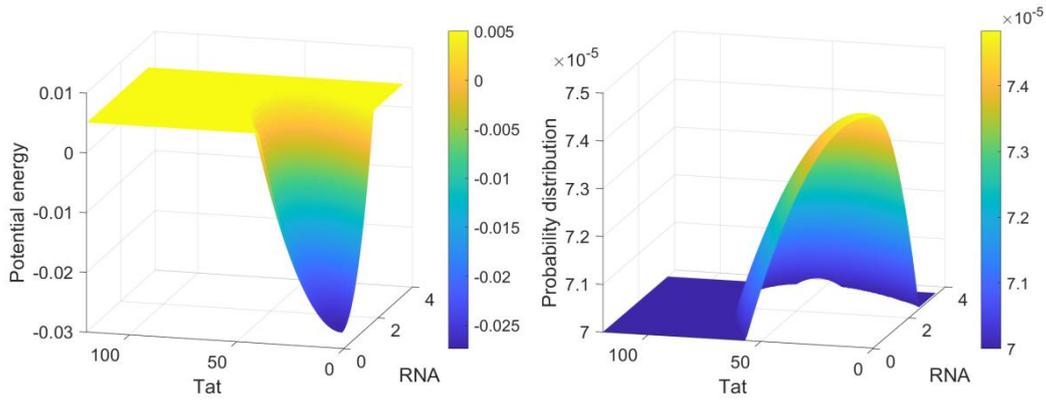

**Fig. S23  Potential energy function and probability distribution diagram when** $K_2 = 10^{-3}$

f．调整参数 $C$ ，出现只有单潜伏态的情形（在此组参数下无法得到单激活态的情形）（取 Tat=50，RNA=1.5 时的势能为零点，并且由于只关注稳定点附近的性质，图中未展现出 $U > 5.0\times10^{-3}$ ，$P < 7\times10^{-5}$ 时其余地方的变化）

Ⅰ：取 $C = 50$ 个时，只有潜伏态（图 S24）

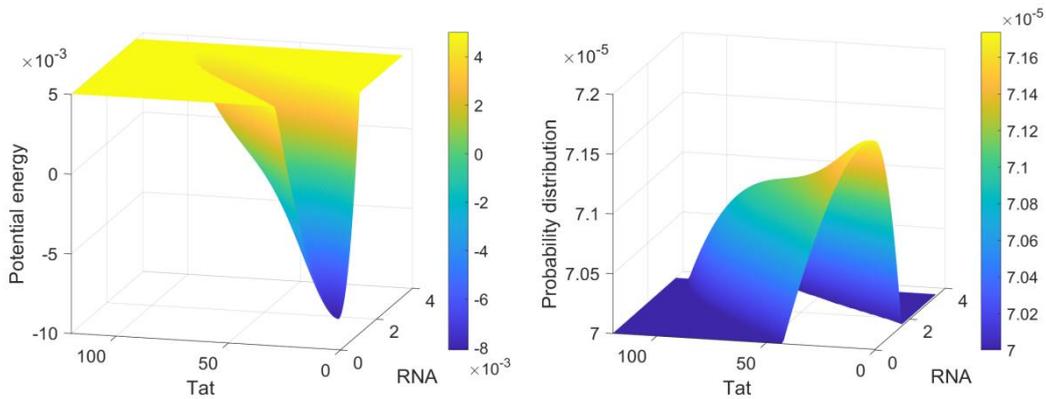

**Fig. S24  Potential energy function and probability distribution diagram when** $C = 50$

g．调整参数 $r$ ，出现只有单激活态和单潜伏态的情形（取 Tat=50，RNA=1.5 时的势能为零点，并且由于只关注稳定点附近的性质，图中未展现出 $U > 5.0\times10^{-3}$ ，$P < 7\times10^{-5}$ 时其余地方的变化）

Ⅰ：取 $r = 150/s$ 时，只有激活态（图 S25）

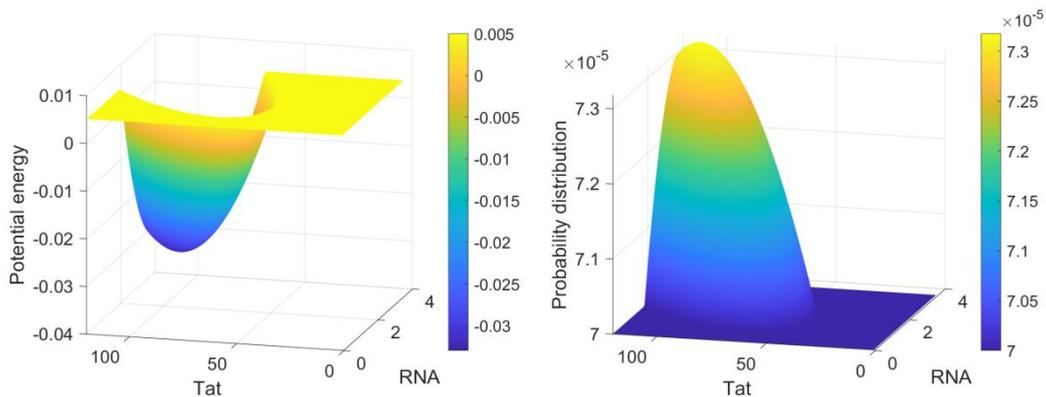

**Fig. S25  Potential energy function and probability distribution diagram when** $r = 150$

Ⅱ：取 $r = 10/s$ 时，只有潜伏态（图 S26）

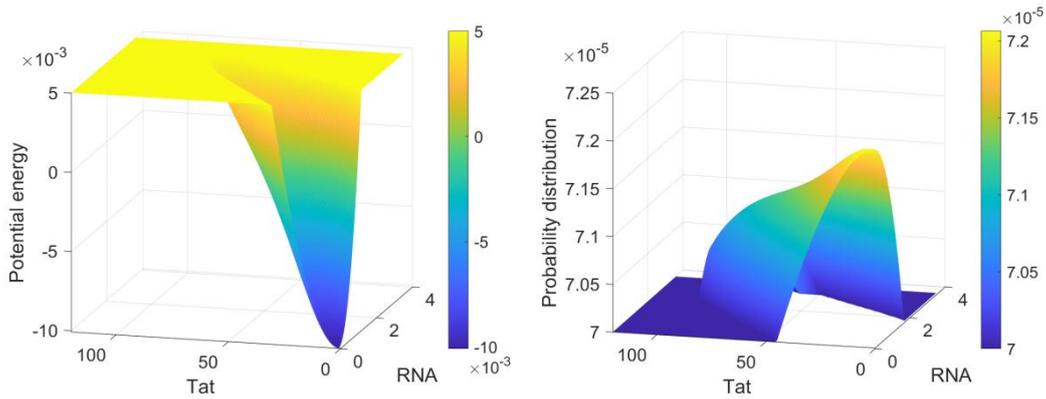

**Fig. S26** Potential energy function and probability distribution diagram when $r=10$

h. 调整参数 $\delta_t$，出现只有单激活态和单潜伏态的情形（取 Tat=280，RNA=3.5 时的势能为零点）

Ⅰ：取 $\delta_t=10^{-5}/s$ 时，只有激活态（由于只关注稳定点附近的性质，图中未展现出 $U>5.0\times10^{-3}, P<6\times10^{-5}$ 时其余地方的变化）（图 S27）

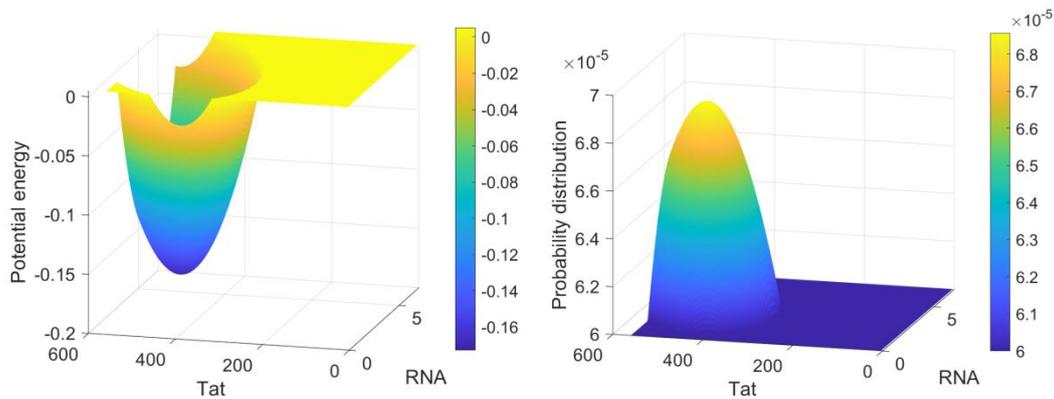

**Fig. S27** Potential energy function and probability distribution diagram when $\delta_t=10^{-5}$

Ⅱ：取 $\delta_t=6\times10^{-5}$ 时，只有潜伏态（由于只关注稳定点附近的性质，图中未展现出 $U>-1.0\times10^{-1}, P<6.3\times10^{-5}$ 时其余地方的变化）（图 S28）

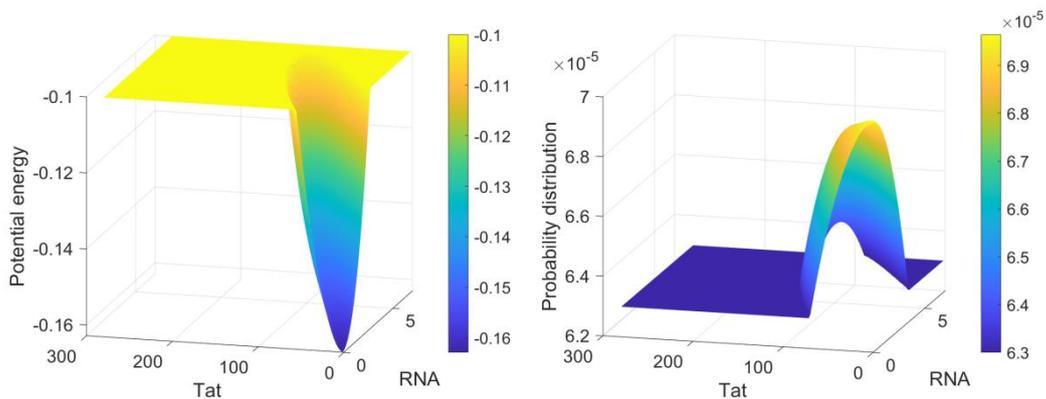

**Fig. S28** Potential energy function and probability distribution diagram when $\delta_t=6\times10^{-5}$

i. 调整参数 $\delta_r$，出现只有单激活态和单潜伏态的情形（取 Tat=280，RNA=10.5 时的势能为零点，并且由于只关注稳定点附近的性质，图中未展现出 $U>5.0\times10^{-3}, P<7\times10^{-5}$ 时其余地方的变化）

Ⅰ：取 $\delta_r=10^{-5}/s$ 时，只有激活态（图 S29）

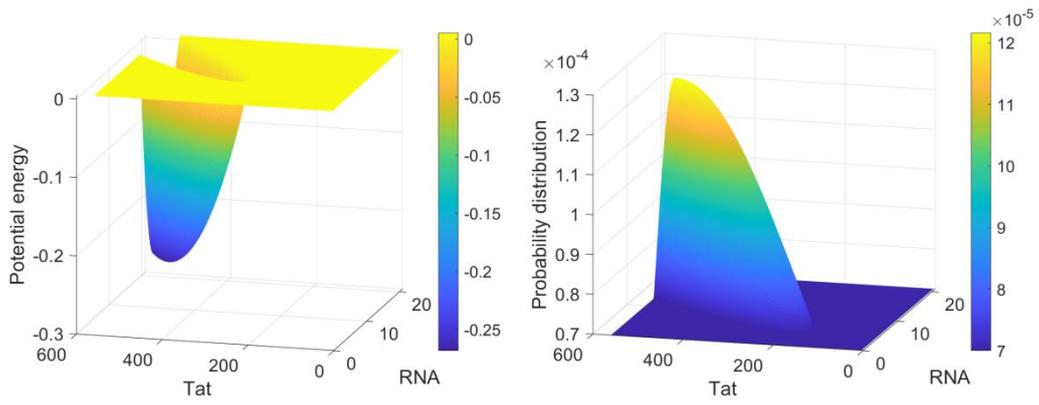

**Fig. S29** Potential energy function and probability distribution diagram when $\delta_r = 10^{-5}$

Ⅱ：取 $\delta_r = 6\times10^{-5}$ 时，只有潜伏态（图 S30）

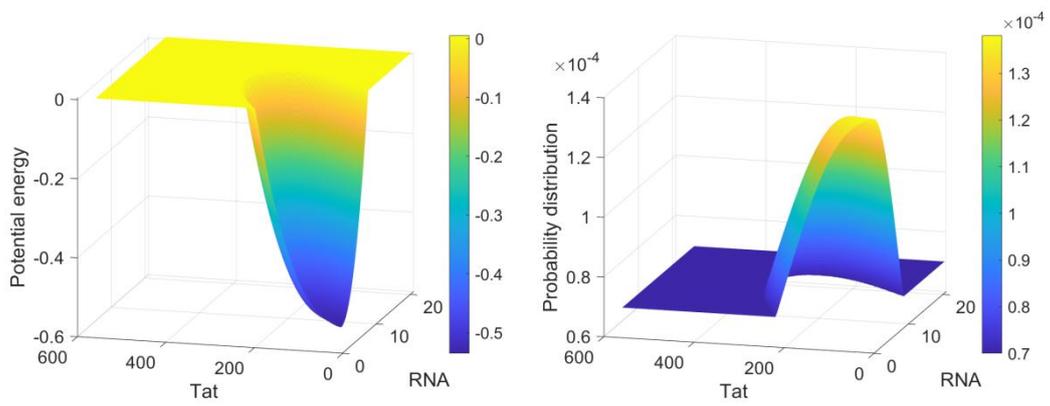

**Fig. S30** Potential energy function and probability distribution diagram when $\delta_r = 6\times10^{-5}$